\documentclass[twocolumn]{aastex631}

\usepackage{tablefootnote}
\graphicspath{{./}{figures/}}

\begin{document}

\title{Elemental Abundances of the Super-Neptune WASP-107b's Host Star Using High-resolution, Near-infrared Spectroscopy}

\author{Neda Hejazi}
\affil{Department of Physics and Astronomy, University of Kansas, Lawrence,  KS 66045, USA }
\affil{Department of Physics and Astronomy, Georgia State University, Atlanta, GA 30303, USA}

\author{Ian J. M. Crossfield}
\affil{Department of Physics and Astronomy, University of Kansas, Lawrence,  KS 66045, USA }

\author{Thomas Nordlander}
\affil{Research School of Astronomy \& Astrophysics, Australian National University, Canberra, ACT 2611, Australia}
\affil{The ARC Centre of Excellence for All Sky Astrophysics in 3 Dimensions, Canberra, ACT 2611, Australia}

\author{Megan Mansfield}
\affil{Department of Astronomy and Steward Observatory, University of Arizona, Tucson, AZ 85719, USA }

\author{Diogo Souto}
\affil{Departamento de F\'isica, Universidade Federal de Sergipe, Av. Marechal Rondon, S/N, 49000-000 S$\tilde{a}$o Crist{\'o}v$\tilde{a}$o, SE, Brazil}

\author{Emilio Marfil}
\affil{Instituto de Astrof\'isica de Canarias, E-38205 La Laguna, Tenerife, Spain}
\affil{Departamento de Astrof\'isica, Universidad de La Laguna, E-38206 La Laguna, Tenerife, Spain}

\author{David R. Coria}
\affil{Department of Physics and Astronomy, University of Kansas, Lawrence,  KS 66045, USA }

\author{Jonathan Brande}
\affil{Department of Physics and Astronomy, University of Kansas, Lawrence,  KS 66045, USA }

\author{Alex S. Polanski}
\affil{Department of Physics and Astronomy, University of Kansas, Lawrence,  KS 66045, USA }

\author{Joseph E. Hand}
\affil{Department of Physics and Astronomy, University of Kansas, Lawrence,  KS 66045, USA }

\author{Kate F. Wienke}
\affil{Department of Physics and Astronomy, University of Kansas, Lawrence,  KS 66045, USA }

\begin{abstract}
We present the first elemental abundance measurements of the K dwarf (K7V) exoplanet-host star WASP-107 using high-resolution (R $\simeq$ 45,000), near-infrared (H- and K-band) spectra taken from Gemini-S/IGRINS. We use the previously determined physical parameters of the star from the literature and infer the abundances of 15 elements -- C, N, O, Na, Mg, Al, Si, K, Ca, Ti, V, Cr, Mn, Fe, and Ni, all with precision $<$ 0.1 dex -- based on model fitting using MARCS model atmospheres and the spectral synthesis code Turbospectrum. Our results show near-solar abundances and a carbon-to-oxygen ratio (C/O)  of  0.50 $\pm$ 0.10, consistent with the solar value of 0.54 $\pm$ 0.09. The orbiting planet, WASP-107b, is a super Neptune with a mass in the Neptune regime (= 1.8 \textit{M$_\textrm{\footnotesize{Nep}}$}) and a radius close to Jupiter’s (= 0.94 \textit{R$_\textrm{\footnotesize{Jup}}$}). This planet is also being targeted by four JWST Cycle 1 programs in transit and eclipse, which should provide highly  precise measurements of atmospheric abundances. This will enable us to properly compare the planetary and stellar chemical abundances, which is essential in understanding the formation mechanisms, internal structure, and chemical composition of exoplanets. Our study is a proof-of-concept that will pave the way for such measurements to be made for all JWST's cooler exoplanet-host stars.
\end{abstract}

\section{Introduction} 
Since the detection of the first planet orbiting a main-sequence star other than the Sun in the 1990s (Mayor \& Queloz 1995), there have been a number of studies comparing the properties of host stars and their planets, in particular their chemical compositions. A host star and its planet are believed to originate from the same molecular cloud, and there has likely been a mutual influence between the two components since their formation. The properties of the host star have a strong impact  on the protoplanetary disk where the planet is formed (e.g., Dorn et al. 2015). Inversely, the accretion of planetary material into the star  by post-formation events such as  planet engulfment may implant the chemical signature of the planet in the atmosphere of the parent star (e.g. Pinsonneault et al. 2001;  Oh et al. 2018; Liu et al. 2018; Ram{\'i}rez et al. 2019; Nagar et al.  2019; Spina et al. 2021; Bonsor et al. 2021; Xu \& Bonsor 2021; Putirka \& Xu 2021).  As a result, the detailed chemical composition of the host star provides fundamental clues on the condition of the protoplanetary disk and the subsequent planetary formation and evolution, as well as the characteristics and habitability of exoplanets.  

The most well-known observational evidence for the chemical link between host stars and their planets is the effect of the host-star metallicity on the planet occurrence rate. Studies have shown that the occurrence rate of giant, close-in ( $<$ 1 au) planets is enhanced  around stars with higher metallicity (e.g., Gonzalez 1997; Heiter \& Luck 2003; Santos et al. 2004; Fischer \& Valenti 2005; Johnson et al. 2010; Mortier et al. 2013), although the detection rate enhancement decreases with decreasing planetary mass and radius (Buchhave et al. 2014; Buchhave \& Latham 2015; Schlaufman 2015; Wang \& Fischer 2015; Mulders et al. 2016; Zhu et al. 2016; Winn et al. 2017; Wilson et al. 2018; Petigura et al. 2018). A similar correlation also exists between the abundances of individual refractory elements (rather than overall metallicity) such as  Mg, Si, Al,  and Ti  and the planet occurrence rate. Based on their statistical method,  Brugamyer et al. (2011) determined a 99{\%} probability that planet detection rate depends on the silicon abundance of the parent star, over and above the observed planet–metallicity correlation. However,  they did not find any such trend for oxygen, i.e.,  the most important volatile element after hydrogen and helium in giant planets (mainly from the accretion of  water ice  beyond the ice line\footnote{The snow line, also known as the  ice line, is the distance in the protoplanetary disk from the center of the disk where it is cold enough for volatile compounds such as water, ammonia, methane, carbon dioxide, and carbon monoxide to condense into solid ice grains.} of the disk, and to a lesser degree, via the oxides of Si, Mg, Ca, and Al). Adibekyan et al. (2012a) also found that the abundance ratio of these refractory elements to iron ([X/Fe]) for giant planet-host stars are systematically higher than those in their comparison sample without detected planets at metallicities [M/H] $<$ $-$0.1 $\pm$ 0.1 dex. Simultaneously, the authors showed that those stars with Neptune-like planets have higher [Ti/Fe] (for [M/H] $<$ $-$0.2 dex), [Si/Fe] and [Al/Fe] (for [M/H] $<$ +0.0 dex), and also [Mg/Fe] (over the entire metallicity range), as compared to non-host counterparts in the comparison sample.

All these correlations can point toward the core accretion model  (e.g., Pollack et al. 1996; Mordasini et al. 2009) as a dominant mechanism for planetary formation. A more metal-rich host star indicates a more solid-rich  protoplanetary disk  (i.e., with a higher solid surface density), which allows the giant planet core to grow from planetesimals more efficiently, and then accrete substantial amount of gas more quickly before the disk dissipates. Note however that these trends have  been mostly examined  for FGK dwarfs, and  the correlation between the chemical composition of  planet-host M dwarfs and the occurrence rate of  orbiting planets is not clear yet. More detailed abundance measurements of M dwarfs together with  the growing number of detected planets around these low-mass stars will reveal the underlying chemical relationship between the two components.

In addition to individual elemental abundances (or their ratios to the iron abundance), the abundance ratio of  volatile elements such as C/O  can shed light on the location in the protoplanetary disk where the planet was formed. The stellar C/O ratio determines the H$_\textrm{\footnotesize{2}}$O, CO, and CO$_\textrm{\footnotesize{2}}$ ice lines in the disk, and can be used to estimate the location of planet formation  when compared to the planetary C/O ratio ({\"Oberg et al. 2011}). A planet having a sub-stellar value of C/O is likely to have a formation location within the H$_\textrm{\footnotesize{2}}$O ice line, and inversely, a planet with a super-stellar C/O value  is likely to have a formation location beyond the H$_\textrm{\footnotesize{2}}$O ice line, and has then  migrated inwards to its current region (see also Reggiani et al. 2022).

The C/O ratio can also place constraints on the planetary   mineralogy (e.g., Delgado-Mena et al. 2021 and references therein). The amount of carbides and silicates formed in planets is specified by the C/O ratio of the parent stars (Larimer 1975; Bond et al. 2010). For example, the stellar C/O can determine if the planetary composition is dominated by carbides or silicates: for high values of C/O ($>$0.8), which indicate carbon-rich systems, Si is more likely to combine with C to form carbides, while for low values of C/O ($<$ 0.8), Si is more expected to combine with O to form silicates, which are the building blocks of rock-forming materials. In low C/O regimes, the type  and distribution of silicates is  governed by the abundance ratio of   refractory elements such as Mg/Si (Thiabaud et al. 2015; Brewer \& Fischer 2017).

Another planet-star chemical connection is related to  the volatile-to-refractory abundance ratios of host stars that have been shown to be correlated with the residual metal of corresponding planets. Using  22  giant planetary systems (including 24 planets with T$_\textrm{\footnotesize{eq}}$ $<$ 1000 K), Teske et al. (2019) found a potential correlation between  the residual planet metals and the ratio of volatile (C and O) to refractory  (Fe, Si, Mg, and Ni) elements of the host stars, which suggests an interesting relationship that could constrain future formation models  of cool, giant planets.

Despite the various studies in the last twenty years to establish correlations between the chemical properties of planets and their parent stars, and the significant progress that has been made in this regard, there has not been any such investigation using an extensive sample of  low-mass host stars, i.e., ``late-type K and M dwarfs". Due to their intrinsic faintness, the acquisition of high-resolution, high signal-to-noise ratio spectra of these cool stars 
demands the use of large 8m-class telescopes and a significant investment in exposure time. The high-resolution spectroscopy required for elemental abundance measurements of cool host stars has therefore been limited to very small, nearby samples. Moreover,  as a result of many blending molecular lines, the complex structure of cool-star spectra makes their analysis difficult. Accordingly,  there have been only a few studies  with reported detailed elemental abundances of low-mass stars (e.g., Souto et al. 2017, 2018, 2020, and 2022, as well as Abia et al. 2020 and Shan et al. 2021 for a couple of specific elements), as compared to numerous analyses of hotter F, G and mid-to-early K dwarfs (e.g., Jofr\'e et al. 2015; Brewer et al. 2016; Delgado Mena et al. 2021 and references therein; Kolecki \& Wang 2022; Polanski et al. 2022, Recio-Blanco et al. 2022). Given the high planet occurrence rate around low-mass stars, high-resolution observations along with new techniques to precisely measure the chemical composition of these stars are needed to better understand the formation and compositions of exoplanets.

 As a pilot study, we developed a spectroscopic method to infer the elemental abundances of a late-type K dwarf harboring a super-Neptune using high-resolution, near-infrared (NIR) spectroscopy.  The planet  is being targeted by four JWST Cycle 1 Guaranteed Time Observation (GTO) programs\footnote{GTO programs 1185, 1201, 1224, and 1280}. Its transmission spectrum will be measured by all four instruments from $\sim0.6-12 \mu$m, and its emission spectrum will be measured with NIRSpec/G395H. However, such measurements can  expand our knowledge about the exoplanet further if placed in the  context of planet–star formation. As a result, the spectroscopic analysis of the parent star is of the same importance as the JWST data and the imminent planetary abundance measurements. Some studies have attempted to infer the chemical abundances of hotter JWST planet hosts (Kolecki \& Wang 2022; Polanski et al. 2022). Since roughly half of JWST's exoplanet sample orbit stars with T$_\textrm{\footnotesize{eff}}$ $<$ 4700 K, it is important to extend  these stellar abundance analyses to cooler JWST host stars based on their available high-resolution spectra or ongoing observations\footnote{We are currently assembling the spectra of some JWST's cool host stars using the IGRINS at the Gemini-South observatory as part of program GS-2023A-Q-203 (PI: Ian Crossfield).}. The comparison between planetary and stellar chemical abundances could then help  elucidate planet formation processes as well as the interplay between the initial composition and present-day chemistry of planetary systems.

 The  description of the planetary system selected for our analysis is detailed in the following section. The  observations from which the spectra were assembled and the data reduction method are summarized in Section 3. In Section 4,  the model atmospheres, linelists and spectral synthesis code that were employed in this study are presented. Our model-fit procedure for measuring the atmospheric chemical abundances of our target is outlined in Section 5. The resulting abundances and their estimated uncertainties are shown in Section 6. Lastly, we discuss our results and compare them with those of other stars in the Solar neighborhood in Section 7.

\section{WASP-107 System}

The star-planet system WASP-107 consists of a host star and  two confirmed exoplanets as described below.

\subsection{Host-Star WASP-107}
The host-star  WASP-107 is a nearby, late-type K dwarf with spectral type K7.0 (Dressing et al. 2019), located at a distance of around 64 pc from the Sun (Gaia Early Data Release 3 or EDR3,   Gaia Collaboration et al. 2021). This star is relatively bright (with apparent Gaia magnitude \textit{G} = 11.18) and has a relatively high proper motion ($\mu$ = 97.12 mas/yr).

WASP-107 is a magnetically active star as manifested by its rotational modulation with a period of \textit{P} = 17.5 $\pm$ 1.4 days (Mo{$\breve{\textrm{c}}$}nik et al. 2017). Starspot occultation events have been detected by bumps or  spot-crossing anomalies in the light curves of this star (Anderson et al. 2017; Dai \& Winn 2017, Mo{$\breve{\textrm{c}}$}nik et al. 2017). Since the rotational period of the star is around three times of the orbital period of planet WASP-107b (see Section 2.2), these occultation events are expected to occur every three transits of the planet.  However, such recurrences have not been observed, which can be attributed to  a high spin-orbit misalignment, assuming that large spots persist for at least one full star rotation (Dai \& Winn 2017, Mo{$\breve{\textrm{c}}$}nik et al. 2017). 

For our analysis, we adopted the physical parameters of the host star determined by Piaulet et al. 2021 using Keck/HIRES spectral analysis: effective temperature T$_\textrm{\footnotesize{eff}}$ = 4425 $\pm$ 70 K, metallicity [M/H] = +0.02  $\pm$ 0.09 dex, and surface gravity  log \emph{g} = 4.633 $\pm$ 0.012 dex. Other stellar parameters are presented in Table 2 of that paper.

\subsection{Exoplanet WASP-107b}
WASP-107b is a warm (T$_\textrm{\footnotesize{b}}$ = 780 K), super-puffy ($\rho_\textrm{\footnotesize{b}}$ =  0.134$_\textrm{\footnotesize{$-$0.013}}^\textrm{\footnotesize{+0.015}}$ g cm$^\textrm{\footnotesize{$-$3}}$),  super-Neptune (\textit{M$_\textrm{\footnotesize{b}}$} = 1.8 $\pm$ 0.1 \textit{M$_\textrm{\footnotesize{Nep}}$} or \textit{M$_\textrm{\footnotesize{b}}$} = 30.5 $\pm$ 1.7 \textit{M$_\textrm{\footnotesize{$\oplus$}}$}) that was first detected based on observations made by the WASP-South photometric survey (Anderson et al. 2017). This planet has already been studied through transit lightcurves by the WASP-South  and  K2 mission (Anderson et al. 2017; Dai \& Winn 2017) as well as  via CORALIE (Anderson et al. 2017) and  Keck/HIRES (Piaulet et al. 2021) radial velocity (RV) measurements. The planet orbits a K dwarf (Section 2.1) at a distance of \textit{a$_\textrm{\footnotesize{b}}$} = 0.0566 $\pm$ 0.0017 au from the star with an eccentricity of \textit{e$_\textrm{\footnotesize{b}}$} = 0.06 $\pm$ 0.04 and a period of \textit{P$_\textrm{\footnotesize{b}}$} = 5.72 days.

The extremely low density of  WASP-107b  (see above)  makes it one of the lowest bulk density planets known,  which suggests a H/He envelope mass faction $>$ 85$\%$ (Piaulet et al. 2021). The planet orbits at the upper border of the Neptune desert\footnote{The Neptune desert (Szab\'{o} \& Kiss 2011; Beauge \& Nesvorny 2013; Mazeh et al. 2016) is an observed scarcity of Neptune-sized planets at short orbital periods. Typically this is explained by atmospheric stripping due to strong stellar irradiation (e.g. Lecavelier des Etangs 2007; Beauge \& Nesvorny 2013; Owen \& Lai 2018), although planetary migration processes have also been theorized to have sculpted the desert’s upper boundary (Owen \& Lai 2018, Bailey \& Batygin 2018), with some observational evidence supporting this (Vissapragada et al. 2022).} (Allart et al. 2019), and given its very large envelope mass fraction, it provides an important  target for  planetary formation and evolution theories. Since the planet's core accreted more than 10 \textit{M$_\textrm{\footnotesize{$\oplus$}}$} in gas, it likely formed at a distance of several astronomical units from the star, where the protoplanetary disk was rich in gas, ice and dust particles, and then has undergone inward migration (Piaulet et al. 2021). However, the mechanism and relevant interactions that  have driven such a migration are still unknown.

Due to its large scale-height and low density atmosphere, and also its small, bright host star, WASP-107b is one the most excellent targets for atmosphere characterization.  The transmission spectra of the planet obtained by the Hubble Space Telescope (HST)/WFC3 (Kreidberg et al. 2018; Spake et al. 2018) and CARMENES (Allart et al. 2019) have been analyzed, and some species such as helium and water  have been identified in its atmosphere. WASP-107b is the first planet on which helium was detected by identifying the narrow absorption feature of excited, metastable helium at 10833 {\AA} (Spake et al. 2018). The signature indicates  an excess absorption in the blue part of the line, suggesting that the planet has an extended, eroding atmosphere whose outer layers are being blown away with an escape rate of metastable helium of about 8$\times$10$^{5}$ gs$^{-1}$, and likely has a gaseous,  comet-like tail caused by radiation pressure (Allart et al. 2019).

The transmission spectra  show strong evidence for water  absorption, which is consistent with a solar abundance pattern. On the other hand, the methane abundance is depleted with respect to expectations for a solar composition pattern, which may be due to either an
intrinsically low C/O ratio or disequilibrium chemistry processes that decrease the amount of methane in the observable portions of the planet's atmosphere  (Kreidberg et al. 2018). The amplitude of water absorption is less than what is expected for a clear, cloud-free atmosphere, and an optically-thick condensate layer at high altitudes is required to model the observed water features. It  is important to  mention that all these results were inferred based on the first measurement of the planet's mass (i.e., \textit{M$_\textrm{\footnotesize{b}}$} = 38 $\pm$ 3 \textit{M$_\textrm{\footnotesize{$\oplus$}}$}, Anderson et al. 2017) and the more accurate mass measurement, which affects the surface gravity estimate, motivates a reanalysis of all reported findings of transmission spectroscopy (Piaulet et al. 2021).

We recall  that the near-future analysis of WASP-107b using JWST spectra 
will allow more precise atmosphere characterization and abundance measurements (with precision $<$ 0.2 dex for gas-rich planets,   Greene et al. 2016), leading to better constraints on the composition and the underlying formation mechanism of the planet.

\subsection{Exoplanet WASP-107c}
During the HIRES spectral analysis of WASP-107b, Piaulet et al. (2021) also identified a significant long-period trend on the top of the signal due to the presence of a second exoplanet,  WASP-107c, and they found a  two-planet Bayesian model  to better match the  HIRES data,  accordingly.  Further, the CORALIE data in the RV analysis indicated another proof for the existence of a second planet, resulting in a two-planet Keplerian solution rather than a one-planet best-fit model.  

The inferred RV semi-amplitude of the outer planet from the HIRES and CORALIE  datasets combined (\textit{K$_\textrm{\footnotesize{c}}$} = 9.6$_\textrm{\footnotesize{$-$1.0}}^\textrm{\footnotesize{+1.1}}$ ms$^\textrm{\footnotesize{$-$1}}$) corresponds to a mass of  \textit{M$_\textrm{\footnotesize{c}}$} sin \textit{i} = 115 $\pm$ 13 \textit{M$_\textrm{\footnotesize{$\oplus$}}$}. The two steep rises found in the HIRES  RV data resulted in well-constrained orbital properties, i.e.,  an orbital period of \textit{P$_\textrm{\footnotesize{c}}$} = 2.98 $\pm$ 0.04 yr and an eccentricity  of \textit{e$_\textrm{\footnotesize{c}}$} = 0.28 $\pm$ 0.07, showing a significantly wider and elongated orbit, compared to that of  WASP-107b. This second, more massive companion may have influenced the migration and the orbital obliquity (spin-orbit misalignment, Section 2.1) of planet b. Considering the sky-projected angular separation of WASP-107c, i.e., 26$_\textrm{\footnotesize{$-$5}}^\textrm{\footnotesize{+8}}$ mas, this planet is too close to the host star to be observed using direct imaging. Additional details of planets b and c are found in Table 4 of Piaulet et al. (2021).

\section{Spectroscopic Observations} 
We employed the high-resolution, NIR spectra of the host star WASP-107 observed using the Immersion GRating INfrared Spectrograph (IGRINS, Yuk et al. 2010; Park et al. 2014) at the Gemini-South observatory. IGRINS is a compact, cross-dispersed spectrograph with a high resolving power (R $\simeq$ 45000) that measures the full coverage of the \textit{H} and \textit{K} bands (1.45-2.45 $\mu$m, except a small gap of about 100 {\AA} between the two bands) simultaneously in a single exposure. IGRINS utilizes a silicon immersion echelle grating and two 2K$\times$2K infrared detectors that allow the spectrograph to obtain spectra at high resolutions in both bands. 

The data were taken on UT 2021-04-19 as part of program GS-2021A-LP-107 (PI: Megan Mansfield), which aimed to conduct transmission spectroscopy of WASP-107b. For our analysis we selected 25 exposures acquired outside of transit,  each with an integration time of 78 s. Specifically, the reduced data were taken from the Raw \& Reduced IGRINS Spectral Archive\footnote{https://igrinscontact.github.io/RRISA{\_}reduced/} (Sawczynec et al. 2022). All spectra in the archive have been reduced using  the IGRINS Pipeline Package (PLP, Lee et al. 2017)\footnote{https://github.com/igrins/plp/tree/v2.1-alpha.3}. To transform the raw data to  final echelle multi order  spectra in the \textit{H} (23 orders) and the \textit{K} (21 orders) bands, the PLP performs a number of reduction processes (flat fielding, background removal, order extraction, distortion correction, and wavelength calibration) and corrects for telluric absorption lines using  telluric standard stars that are normally A0V stars. Telluric standards are usually divided by a model of the Vega spectrum to remove the prominent hydrogen absorption lines in A0V stars. Wavelength solutions are obtained in multiple steps as follows.  Wavelength calibration is primarily derived from an initial guess based on historical wavelength solutions. The resulting calibration is then refined using sky OH emission lines in a 300-second SKY frame taken each night on the telescope. This  solution is further refined using telluric absorption features in the standard star at wavelengths greater than 2.1 $\mu$m.

 The reduced spectra were then combined using the \texttt{combspec} utility that is part of the \texttt{SpeXTool} package (Cushing et al. 2004).  The resulting  stacked spectrum spans wavelengths from 14659--18165 {\AA}  (\textit{H} band) and  19274--24841 {\AA} (\textit{K} band) with essentially no gaps in either band, and  with  median S/N of 517 and 435, respectively.

\section{Model Atmospheres, Line Data, and Spectral Synthesis}
The MARCS model atmospheres (Gustafsson et al. 2008) were used in the present synthesis analysis. These are one-dimensional hydrostatic models, which are computed under plane-parallel geometry and assuming local thermodynamic equilibrium (LTE), along with standard mixing-length theory for convection.  Despite all these approximations, they have been successfully used in a variety of studies ranging from individual stars in our Galaxy to the   stellar populations  and evolution of external galaxies (e.g., An et al. 2009; Davies et al. 2010; Lindgren et al. 2016; Souto et al. 2017, 2018, and 2022; Bensby et al. 2021; Recio-Blanco et al. 2022). Although an extensive model grid can be found in the MARCS website\footnote{https://marcs.astro.uu.se/index.php}, we further used the interpolation routine developed by Thomas Masseron (which is also available in the MARCS website\footnote{https://marcs.astro.uu.se/software.php}) to interpolate the model with physical parameters the same as those of our target star (Section 3.3). 

We employed the atomic line data  taken from the Vienna Atomic Line Database (VALD, Piskunov et al. 1995; Kupka et al. 2000; Heiter et al. 2008; Ryabchikova et al. 2015), a collection of atomic and molecular transition parameters for astronomical purposes, which has been used in various studies of cool stars (e.g., Lindgren et al. 2016; Pavlenko 2017; Reiners et al. 2018; Woolf \& Wallerstein 2020; Muirhead et al. 2020; Delgado Mena et al. 2021; Marfil et al. 2021; Olander et al. 2021; Cristofari et al. 2022; Ishikawa et al. 2020 and 2022). The molecular line data have been  assembled from multiple sources, such as VALD (particularly, for TiO lines in the optical region),  the Kurucz (Smithsonian) Atomic and Molecular Database (Kurucz 1995), and the high-resolution transmission molecular absorption database (HITRAN, Rothman 2021). More specifically, we mention the linelist references of the most important molecular bands used in this study as follows: H$_\textrm{\footnotesize{2}}$O (Barber et al. 2006), OH (Goldman 1982), CO (Goorvitch 1994), FeH (Dulick et al. 2003), and CN (Brook et al. 2014; Sneden et al. 2014).  

We generated the required synthetic spectra using the LTE spectral synthesis code Turbospectrum\footnote{http://ascl.net/1205.004} (TS, Alvarez \& Plez 1998; Plez 2012) version v15.1, together with the MARCS models and a selected set of atomic and molecular linelists, assuming  the solar abundances from Grevesse et al. (2007).

\section{Elemental Abundance Analysis}    
As one would expect, the spectra of late-type K dwarfs are substantially similar to those of M dwarfs  with nearly the same complications in spectral analysis. Such spectra are dominated by numerous molecular lines in both optical and NIR regions. Particularly, the H$_\textrm{\footnotesize{2}}$O, OH, FeH,  and CO molecular bands are blended with many atomic lines in the NIR spectral region. As a result, equivalent width analysis to measure individual elemental abundances does not apply to these spectra, and spectral synthesis would provide the best approach to infer the detailed chemical composition of our target. In this  work, we measured the abundances of fifteen elements --  C, N, O, Na, Mg, Al, Si, K, Ca, Ti, V, Cr, Mn, Fe, and Ni --  using an iterative  synthetic spectral fitting in both the \textit{H} and \textit{K} bands.  It should be pointed out that we only used synthetic ``continuum-normalized" spectra (generated by TS+MARCS), and hereafter, we call them ``synthetic spectra" or ``synthetic models" for simplicity. Our method is outlined in the following sections.
 
\subsection{Pre-processing and Radial Velocity Shift}
 The observed spectra underwent  some pre-processes before being used in the fitting routine. We  first divided the spectra in both bands  in  smaller parts of 100--200  {\AA}  wide, and flattened each part  by fitting a low-order (second or third order) polynomial. We then carried out a careful visual inspection over all small parts to exclude spectral anomalies and problematic regions that could be due to bad pixels, instrumental  artifacts, or imperfect data reduction. For ease of spectral fitting analysis, all the remaining segments were combined together to make a single spectrum, spanning from the  \textit{H} to \textit{K} band.

We compared the observed spectrum, whose wavelengths were Doppler shifted, with a good estimate of best-fit synthetic model corresponding to  the star's physical parameters (i.e., T$_\textrm{\footnotesize{eff}}$ = 4425 K,  [M/H] = +0.02 dex, and  log \emph{g} = 4.633 dex) and  assuming a microturbulence parameter of $\xi$ = 1.00 km/s along with approximate elemental abundances of  A(X)$_\textrm{\footnotesize{approx}}$ = A(X)$_\textrm{\footnotesize{$\odot$}}$ + [M/H], where A(X)$_\textrm{\footnotesize{$\odot$}}$ is the  solar abundance of element X. We examined different radial velocity (RV) values and found a best-fit of 105 $\pm$ 1 km/s for the target\footnote{This velocity was inferred from pure  spectral synthetic fitting, and no  radial velocity calibration or corrections, for example including the motion of the Earth relative to the Sun, were considered. More importantly, there is an offset between the IGRINS wavelengths calibrated in vacuum and MARCS synthetic spectra calibrated in air, and no vacuum-to-air wavelength conversion has been made for the above radial velocity. As a result, it does not represent the true radial velocity of  the star (e.g., 13.74 km/s from Gaia Collaboration 2018).}. The wavelengths of the target's spectrum were then shifted according to this best-fit RV value before passing through the fitting process.

\subsection{$\chi$$^\textrm{\footnotesize{2}}$ Minimization and Continuum/Pseudo-continuum Placement}
The model fitting was performed by a $\chi$$^\textrm{\footnotesize{2}}$ minimization (including the random error of the observed flux at each wavelength) over an interval (fitting window or $\chi$$^\textrm{\footnotesize{2}}$ window) around the core of the lines of interest individually. While fitting, the synthetic  spectra were convolved using a Gaussian broadening kernel at the observed spectral resolution, and were then interpolated at the shifted, observed wavelengths. Subsequently, the continuum/pseudo-continuum placement was determined using a procedure similar to that described in Santos-Peral et al. (2020).  This is of great importance in the synthetic fitting of cool stars whose pseudo-continuum levels are lower than unity. For this purpose, the continuum/pseudo-continuum regions around  each line, or around a few lines if they are very close to one another, were carefully determined. The observed spectrum was then renormalized relative to a given model spectrum using some data points within these continuum/pseudo-continuum regions. The best such data points  were  selected using a low-order polynomial fit over the residuals, R = O/S, where O is the observed flux and S is the interpolated synthetic flux at each shifted wavelength, followed by a $\sigma$-clipping with three iterations in order of  2$\sigma$, 1.5$\sigma$, and 1$\sigma$.  A final polynomial fit over the residual of the selected data points was obtained, and this fit was then evaluated at all wavelengths around the analyzed line, including both the continuum/pseudo-continuum regions and the $\chi$$^\textrm{\footnotesize{2}}$ window. The renormalized spectrum was determined after dividing the observed flux by this final polynomial-fitted residual. In each $\chi$$^\textrm{\footnotesize{2}}$ minimization run,  the renormalized observed spectrum was compared with  a set of synthetic  models to infer the best-fit solution. Figure 1 shows four different spectral regions around a few spectral atomic and molecular lines used in this analysis. The renormalized observed spectrum (red dots) is compared to the final best-fit model (blue lines, see Section 6). The green dots are the best selected points in the continuum/pseudo-continuum regions used for renormalization and the shaded regions are the $\chi$$^\textrm{\footnotesize{2}}$ window.

\subsection{Spectral Line Identification}
Using the spectral line lists, we identified the atomic lines that were strong enough to be distinguished from the background molecular opacities. We renormalized the observed  spectrum around each spectral line relative to the rough estimate of  the star's best-fit  synthetic model as described in Section 5.1. We then visually compared the resulting spectrum with that model, and removed those lines that were noticeably discrepant from their respective lines in the model spectrum, whether in depth or shape. These differences may be due to spectral noise, artifacts, or the insufficient modeling of atomic lines and/or blended molecular bands. We also found some lines that had no correspondence in the synthetic spectrum, and thus were excluded  from our fitting analysis. These lines might be due to residual telluric lines, or caused by unknown species that are not included in the spectral synthesis, or  might arise from  transitions that are missing in the linelists used in the analysis, and need to be characterized in the future.  The number of selected lines (\textit{N}) for each element, which were used in our elemental abundance measurements, are depicted in the second column of Table 1 (that also shows the final results of this study, see Section 6). The NIR region is dominated by the atomic lines of Fe,  and the molecular lines of  OH (in the \textit{H} band) and CO (in the \textit{K} band), and consequently, the majority of the lines chosen for the abundance analysis correspond to these three species. In general, the atomic lines of carbon and oxygen are too weak, and are mostly blended with the lines of other species in the spectra of cool dwarfs. For this reason, the molecular OH and CO lines are used  to measure the abundance of carbon and oxygen, respectively. As shown in Table 1, we determined only one well-defined line for three elements: K, V, Cr, and Mn.

\subsection{Microturbulent, Macroturbulent, and Rotational Velocity}
Prior to measuring chemical abundances, we determined the microturbulence parameter $\xi$ based on the method described in Souto et al. (2017). If this parameter is not customized, TS calculates the requested synthetic spectrum using the  default value $\xi$ = 1 km/s, which may not represent the best value for the star under analysis. Souto et al. (2017) found that the synthetic spectra showed little sensitivity to the  microturbulent velocity over most spectral lines, except for  the OH lines. They estimated the microturbulent velocity by measuring the oxygen abundances for a number of  OH lines using different values of  $\xi$ ranging from 0.5 to 1.5 km/s, in steps of 0.25 km/s, and then selected the $\xi$ value that  showed the lowest spread in the  abundances. However, their study was limited to the \textit{H} band where only a few CO lines can be identified. In contrast, our IGRINS spectrum covers both the \textit{H} and \textit{K} bands, which provide us with a significantly larger number of strong Fe and CO lines. As seen from Table 1, the number of our selected OH, CO, and Fe lines is statistically large enough to investigate the sensitivity of these species to microturbulent velocity. We measured the abundances of oxygen, carbon, and iron from our selected OH, CO, and Fe lines, respectively, using the  $\chi$$^\textrm{\footnotesize{2}}$ minimization procedure described above. We used the synthetic spectra associated with the star's physical parameters and abundances equal to A(X)$_\textrm{\footnotesize{approx}}$  for all elements other than the analyzed one, and  examined the resulting abundances  for different values of microturbulent velocity ranging from 0.5 to 2.50 km/s, in steps of 0.25 km/s. We then calculated the standard deviation of abundances for each species and for each $\xi$ value, and found that the scatter of abundances inferred  from  CO lines  changes ten times  more than that of abundances inferred from OH and Fe lines over the selected range of the $\xi$ parameter, which indicates the CO lines as the most sensitive to this parameter. Figure 2 presents the variation of the scatter with respect to the $\xi$ parameter, showing  a clear minimum at $\xi$ = 1.25 km/s that we adopted as the best value of microturbulent velocity. This value is consistent with the other $\xi$ values used in various studies of cool stars  usually  between 1 and 2 km/s (e.g., Becker et al. 2008; Tsuji \&  Nakajima 2014; Pavlenko 2017; Olander et al. 2021; Recio-Blanco et al. 2022). 

Apart from $\xi$,  the spectral line broadening due to other parameters such as  macroturbulence velocity ($\zeta$) and rotational velocity (\textit{v}$_\textrm{\footnotesize{rot}}$sin(i)) can also be important in abundance analysis. However, in contrast to  $\xi$, the values of $\zeta$ and \textit{v}$_\textrm{\footnotesize{rot}}$sin(i) cannot be freely chosen as input parameters when generating  synthetic models using TS code. To this end, we took account of  these two  parameters  through  a post-processing convolution using a Gaussian kernel similar to the one used for instrumental broadening. We performed a $\chi$$^\textrm{\footnotesize{2}}$ minimization over the entire spectrum using the roughly estimated best-fit model, i.e., assuming the best-fit parameters (including the above-inferred value of $\xi$) and abundances the same as A(X)$_\textrm{\footnotesize{approx}}$ for all elements, while fine-tuned the window length of the smoothing kernel. The inferred window length was then used for our following analysis.

\subsection{Iterative Synthetic Model Fitting}
We performed an iterative $\chi$$^\textrm{\footnotesize{2}}$ minimization process for each single element individually by varying its abundance while keeping the physical parameters  T$_\textrm{\footnotesize{eff}}$, [M/H], log \emph{g}, and $\xi$ fixed equal to their previously determined values. In each iteration, we changed the elemental abundances within a specific range around zero from $-$0.40 to +0.40 dex in steps of 0.01 dex, and implemented a polynomial fit over the resulting $\chi$$^\textrm{\footnotesize{2}}$ values to find the abundance that minimized the polynomial-fit function. In the first iteration, we changed the abundance of each element while assuming abundances  equal to A(X)$_\textrm{\footnotesize{approx}}$ for the other 14 elements. In the next iteration, we repeated the same procedure, varying the abundance of each  element but using the updated abundances from the results of the first iteration for the other elements. This process was iterated until the abundances are converged ($<$ 0.01 dex) to their final values all at the same time.  Since the resulting abundances from each iteration fell well within the selected range, i.e., [$-$0.4,+0.4] dex, the values outside the range were not examined in the model-fit procedure. The best-fit elemental abundances of the star were obtained by taking average over the abundances of multiple lines (if applicable) for each single element.

 \begin{figure*}
\gridline{\fig{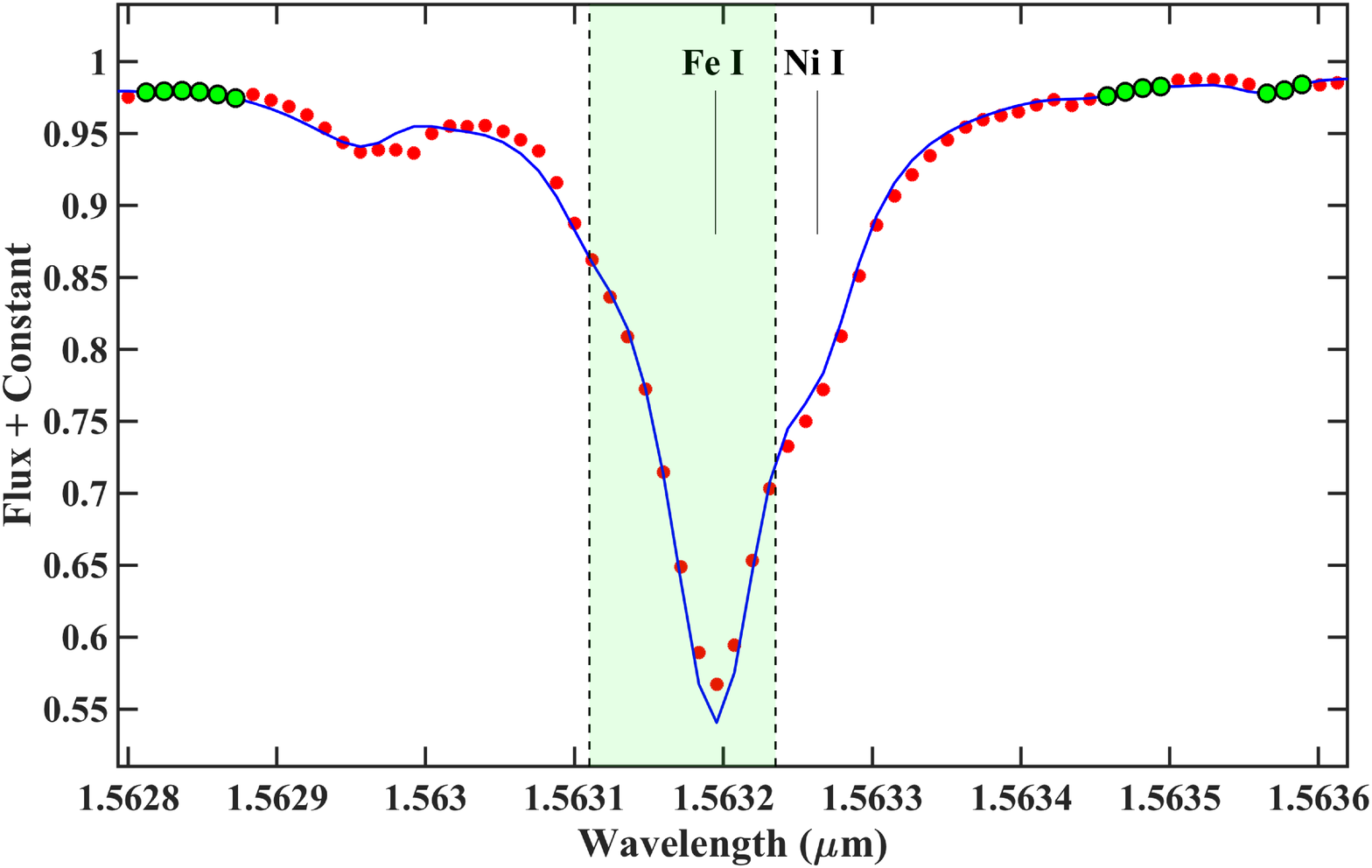}{0.51\textwidth}{}
\hspace{-1.44cm} 
          \fig{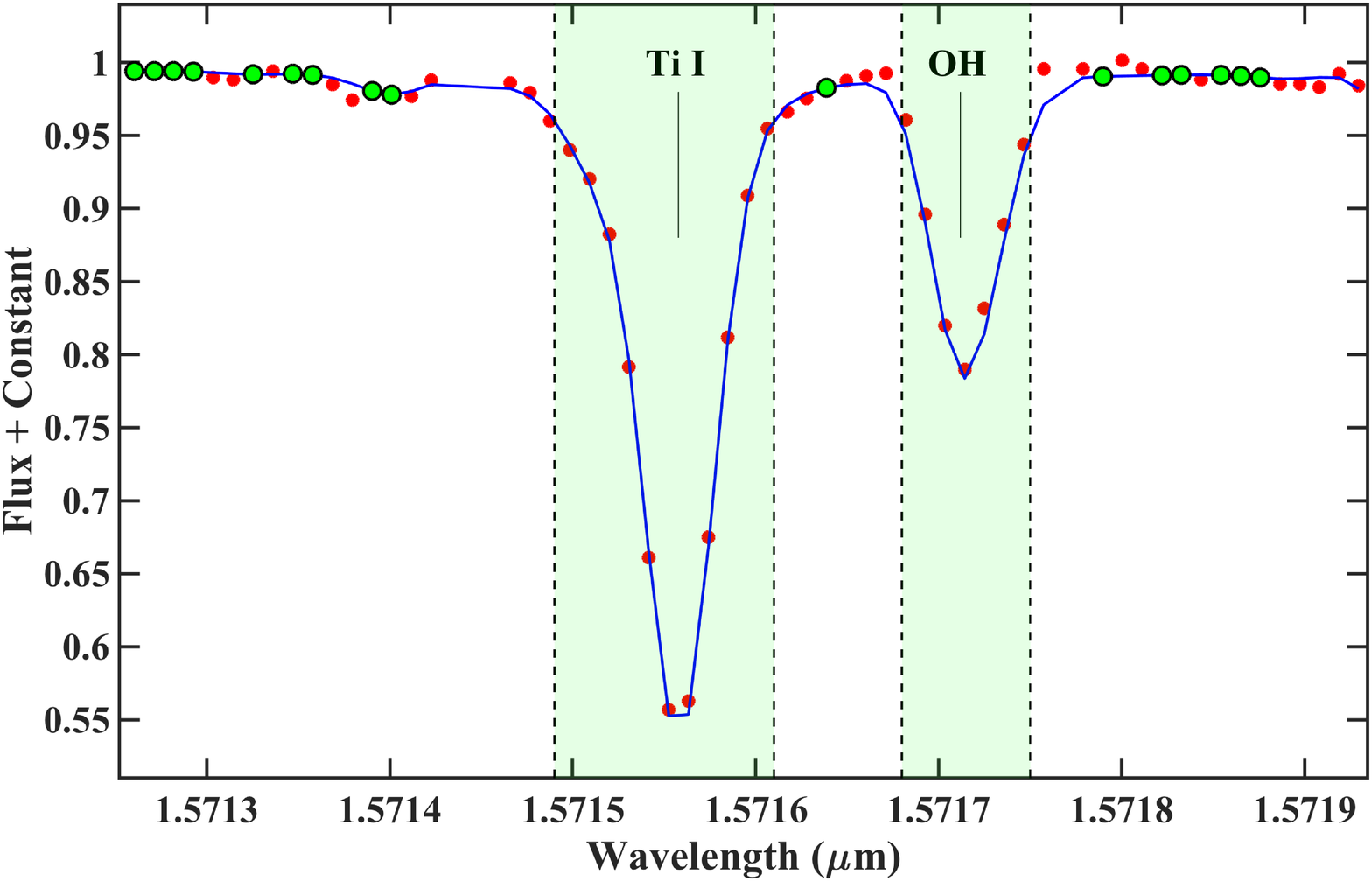}{0.51\textwidth}{}
          }  
          
\gridline{\fig{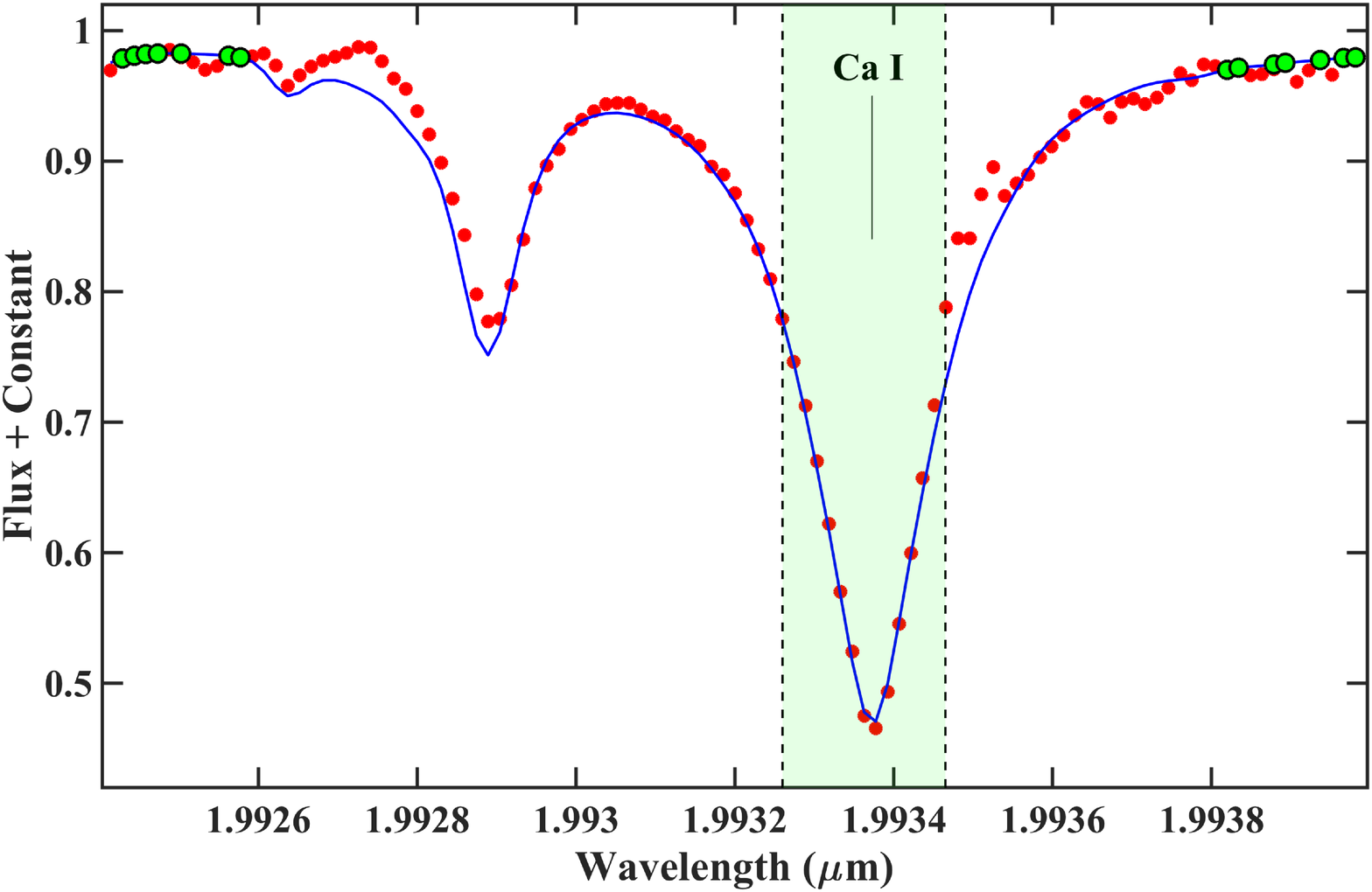}{0.51\textwidth}{}
\hspace{-1.44cm}
          \fig{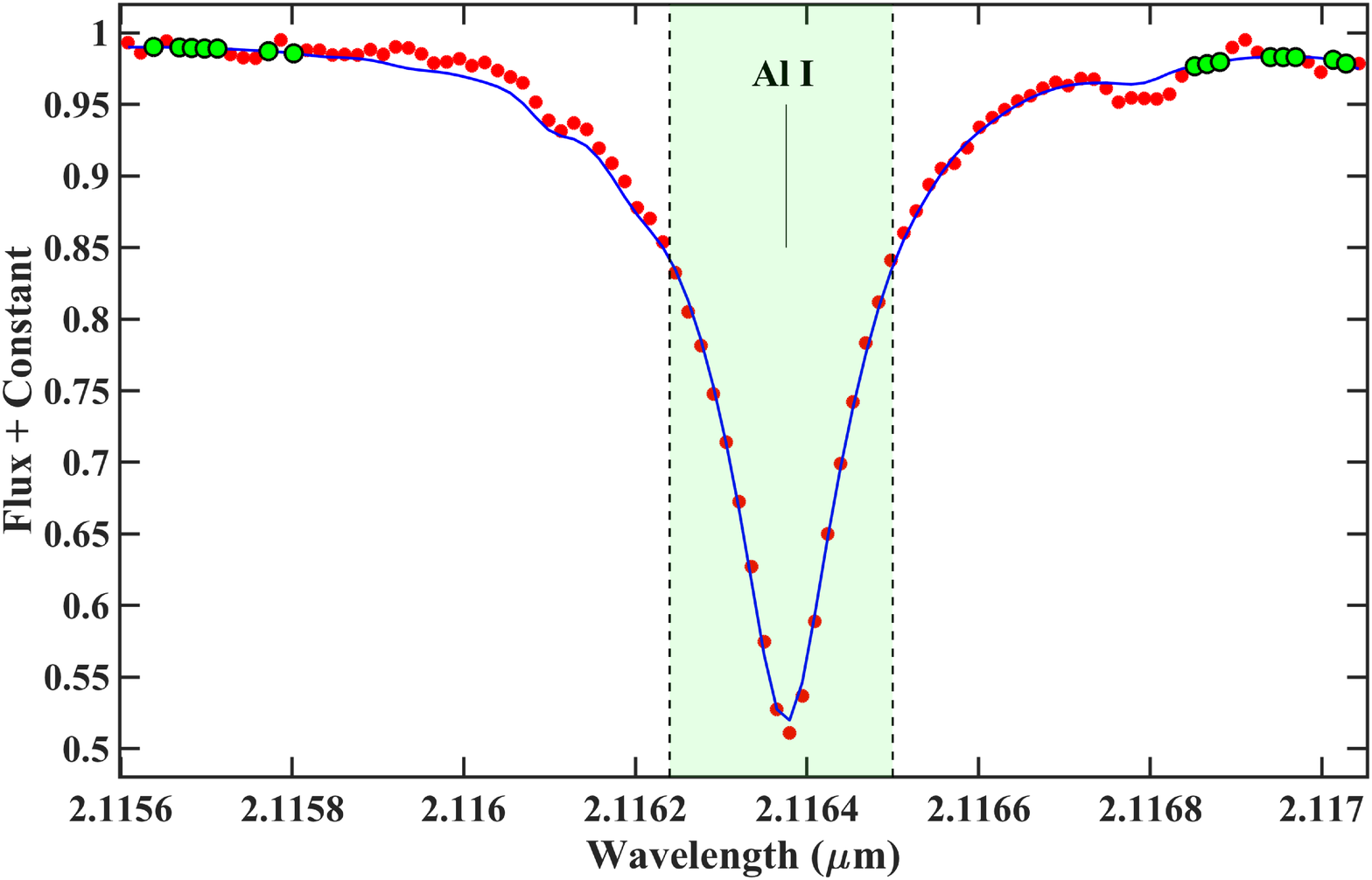}{0.51\textwidth}{}
          }

\gridline{\fig{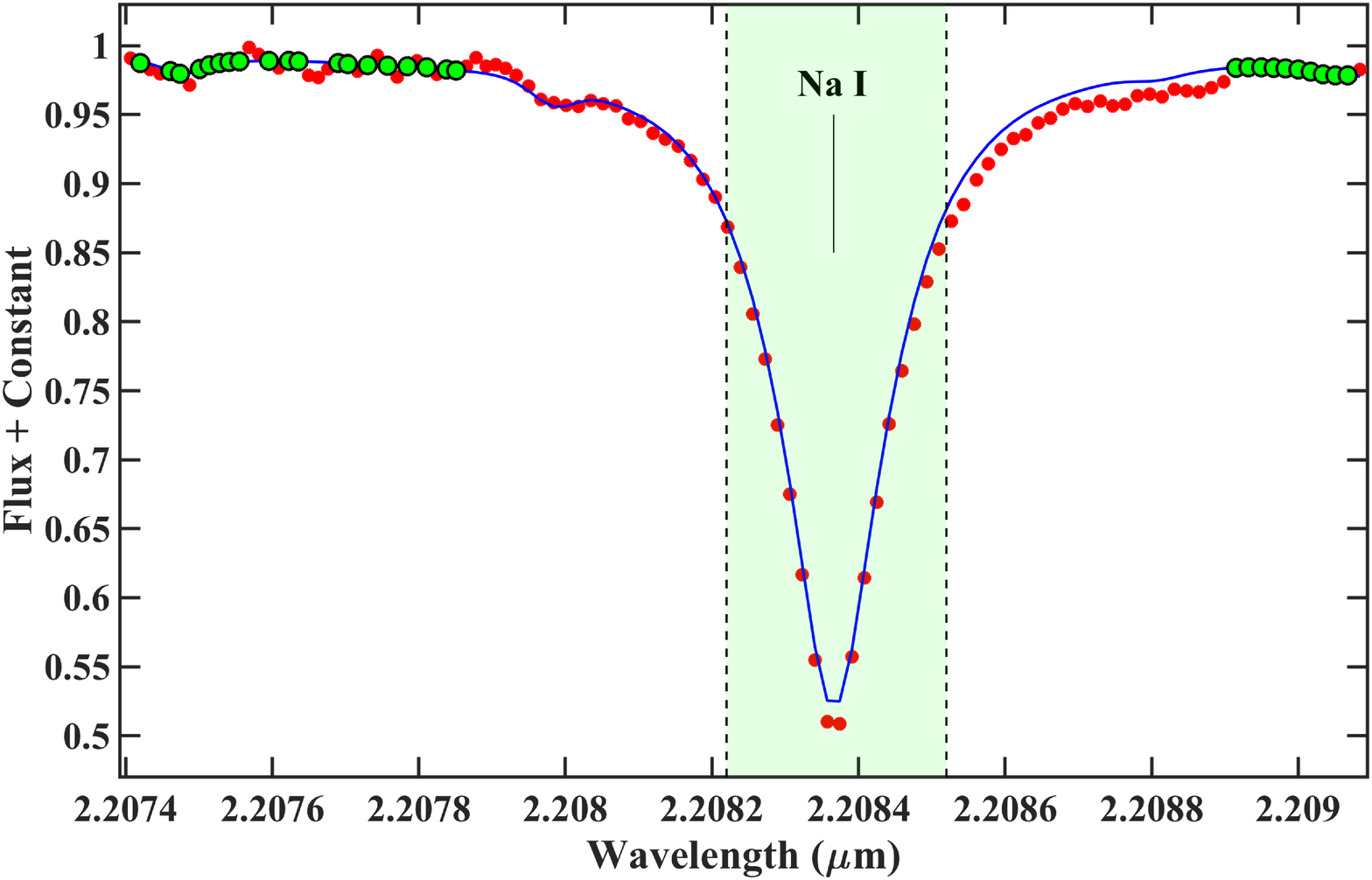}{0.51\textwidth}{}
\hspace{-1.44cm}
          \fig{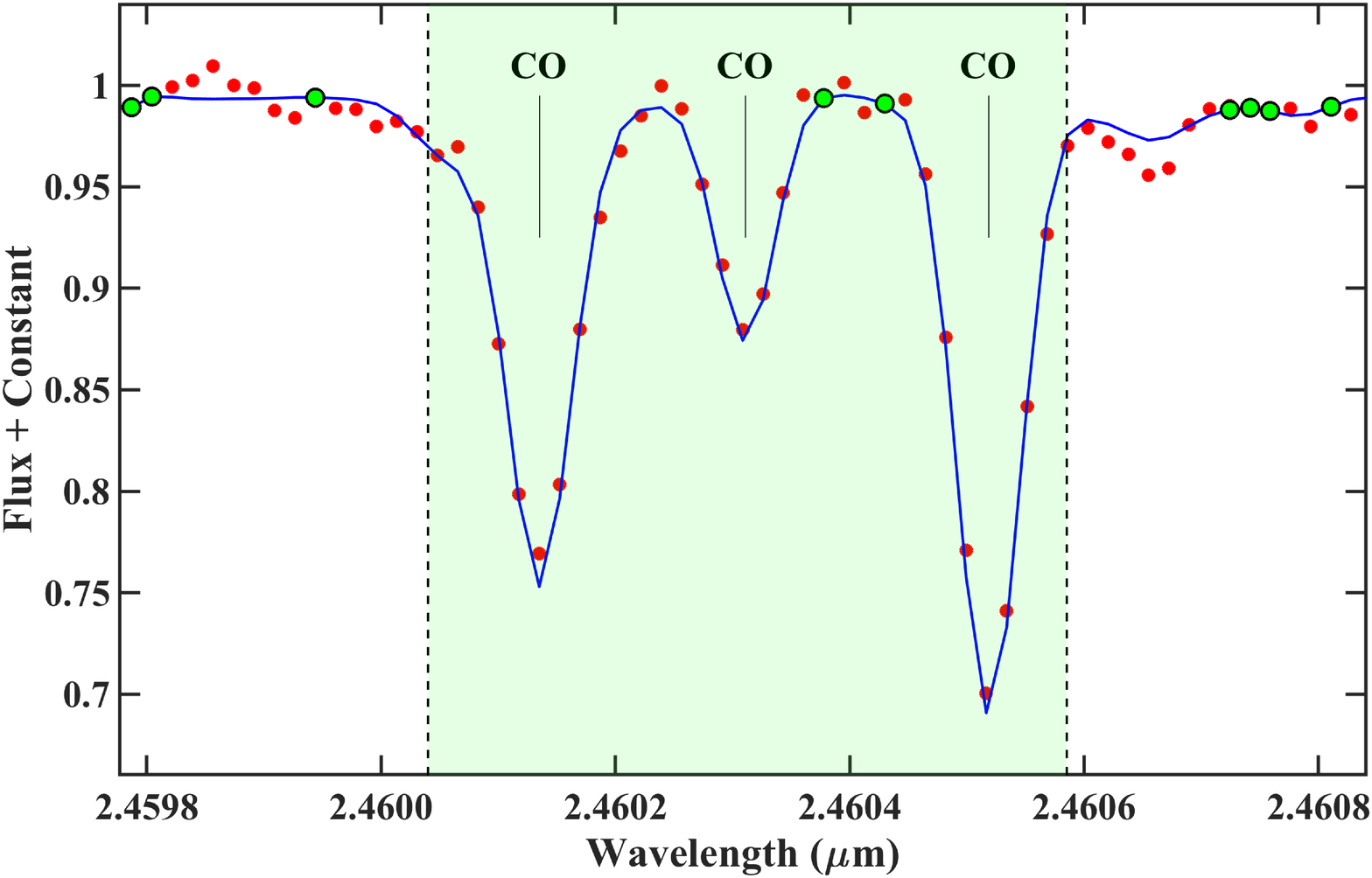}{0.51\textwidth}{}
          }          
\caption{Comparison between the renormalized, observed spectrum (red dots) and the best-fit synthetic model (blue lines) over four different spectral regions. The green dots are the best selected points in the continuum/pseudo-continuum regions used for renormalization and the shaded regions are the $\chi$$^\textrm{\footnotesize{2}}$ windows.}
\end{figure*}

\section{Results}

\subsection{Derived Abundances}
We applied our iterative model-fit procedure to fifteen elements, i.e., C, N, O, Na, Mg, Al, Si, K, Ca, Ti, V, Cr, Mn, Fe, and Ni simultaneously.  The abundances relative to the respective solar values, i.e., [X/H],  and  the absolute abundances, i.e., A(X), where X denotes each of the above-mentioned elements, are listed in the third and fourth columns of Table 1, respectively. Given the near-solar metallicity of the star ([M/H] = +0.02 dex), the individual elemental abundances also have near-solar values. 

As mentioned above, the number of lines associated with CO,  OH, and Fe lines are statistically large enough to present the abundance distribution of these three species, as shown in Figure 3. The top panel shows the distribution of the oxygen abundances with the smallest scatter  (i.e.,  a standard deviation of std = 0.017 dex) compared to the distribution of the carbon (std = 0.041 dex) and iron (std = 0.077 dex) abundances with significant larger dispersions, as shown in  the middle and bottom panels, respectively. The evident consistency between the abundances of different OH lines  indicates the high accuracy of line-dependent quantities such as \textit{gf}-value (the product of the statistical weight and the oscillator strength for a given transition) which are used in the modeling of these lines. On the other hand, the higher line-to-line dispersion in the abundances of carbon and iron as well as  some other elements can be largely due to the uncertainties in the \textit{gf}-values. In addition,  the uncertainties in the continuum placement may also cause discrepancies in the multiple-line abundances of an element. Furthermore,  the inaccuracy of  stellar parameters can raise  significant errors in the abundance of each spectral line (Section 6.2), which can be another reason for such scatters in the measured elemental abundances (Souto et al. 2016 and 2021). Deviations from local thermodynamic equilibrium (LTE) may also influence the derived abundances of particular elements in cool dwarfs (e.g., Olander et al. 2021).

\begin{figure}
\centering
\gridline{\fig{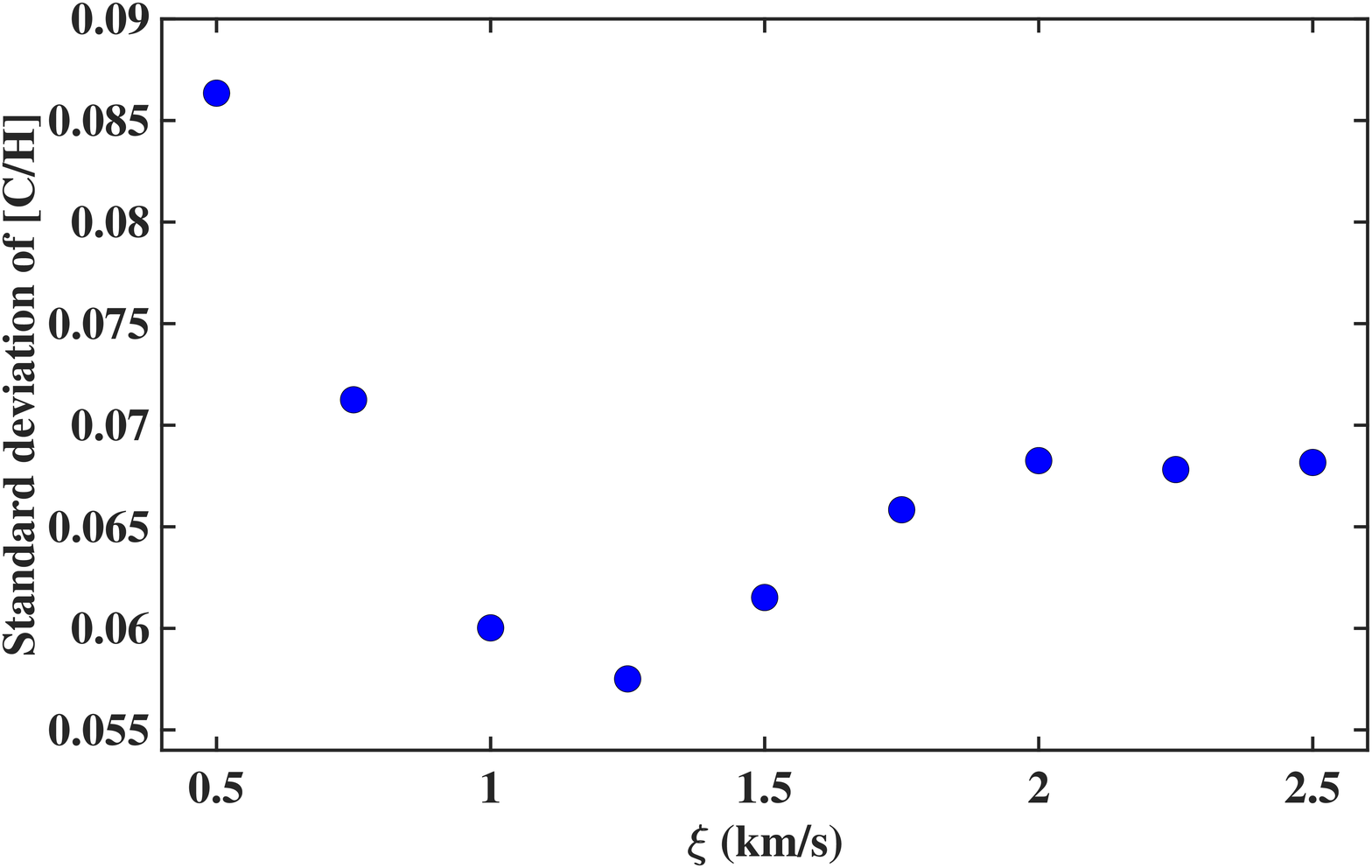}{0.45\textwidth}{}
}  
\caption{Standard deviation of abundances inferred from the selected CO lines for nine different values of the microturbulence parameter $\xi$.}
\end{figure}

\begin{figure}
\centering
\gridline{\fig{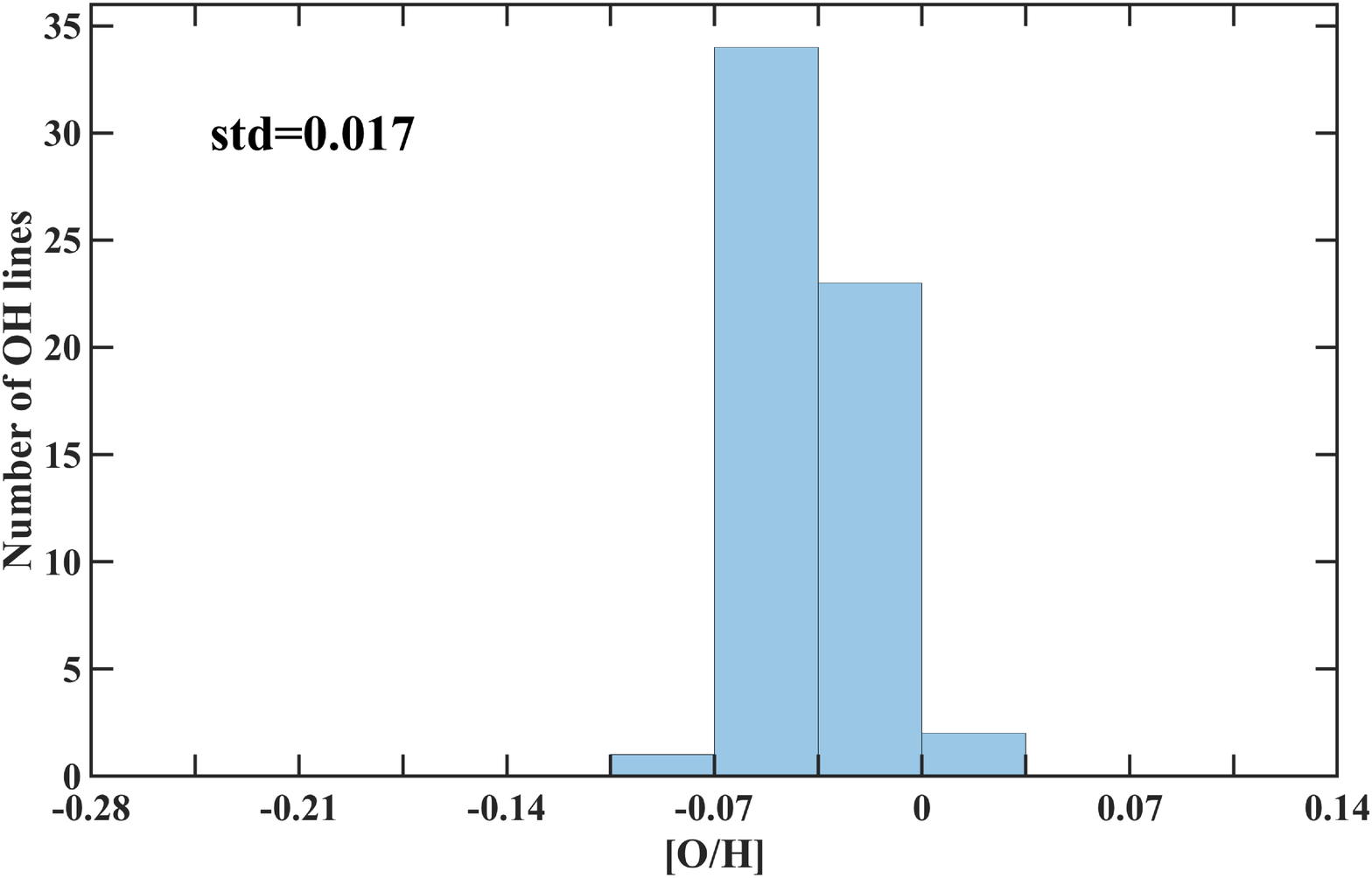}{0.54\textwidth}{}
}  

\vspace{-1.cm}          
\gridline{\fig{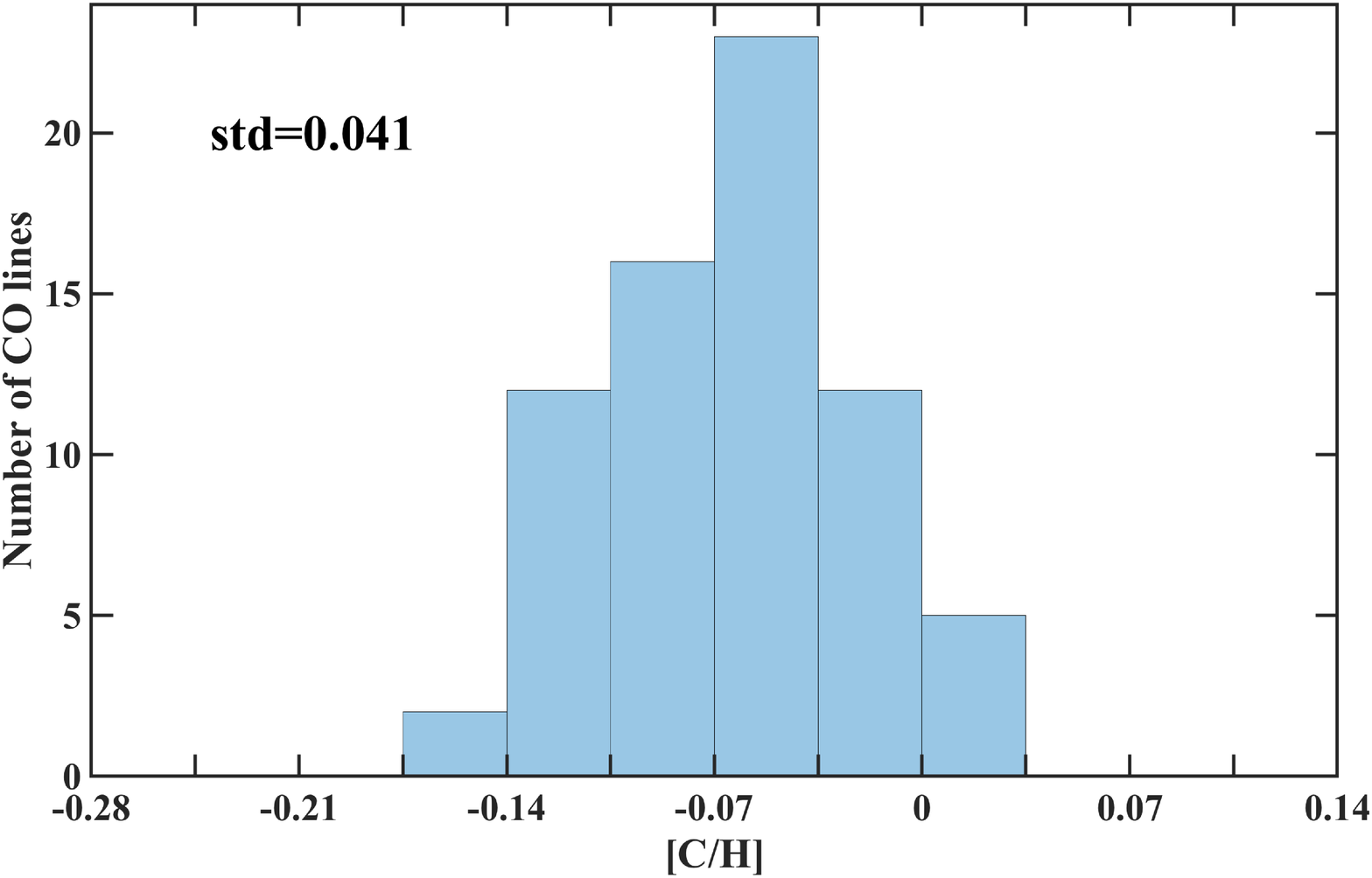}{0.54\textwidth}{}
}

\vspace{-1.cm} 
\gridline{\fig{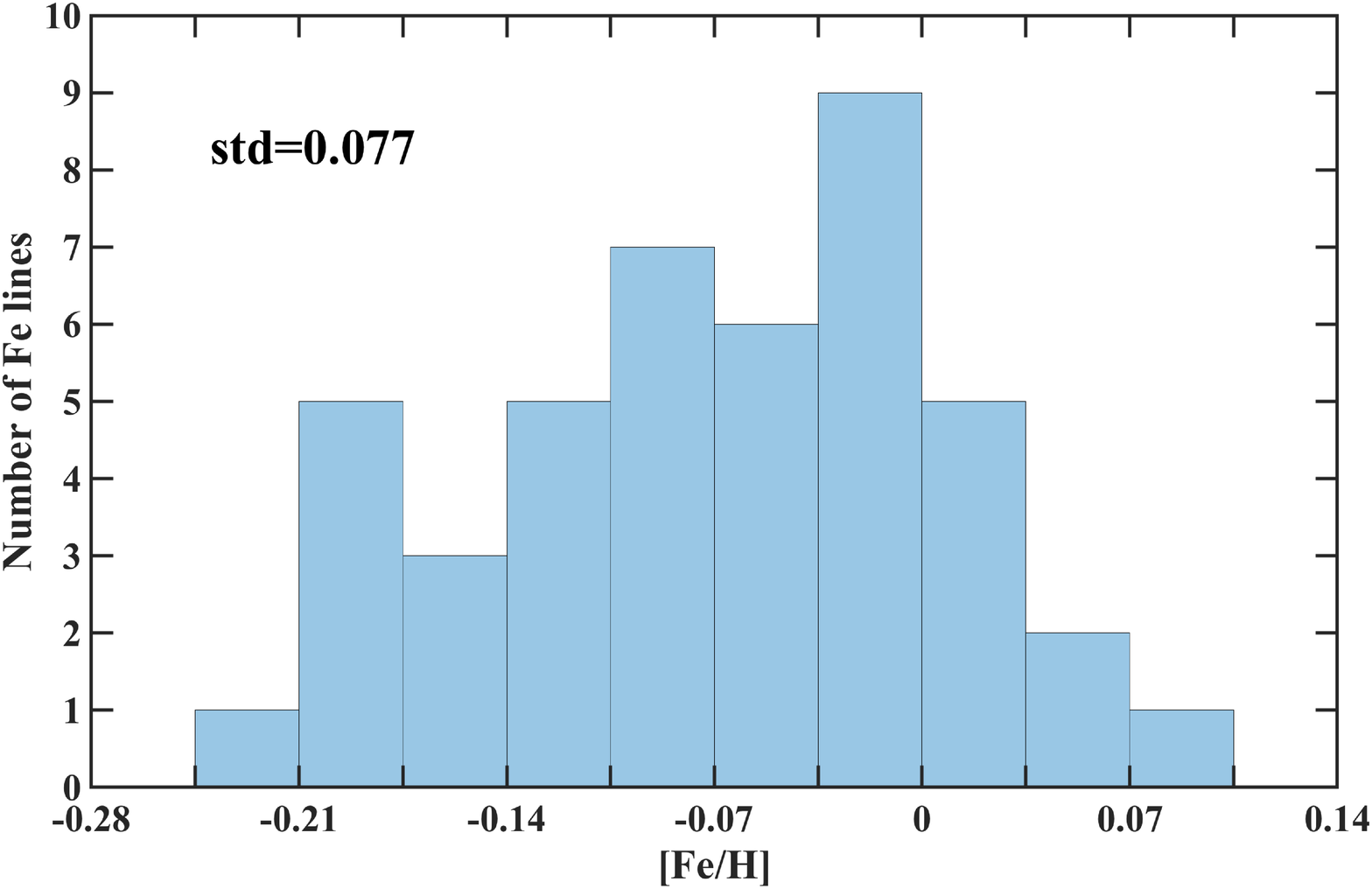}{0.54\textwidth}{}
}
\caption{Abundance distributions of individual lines for oxygen (top), carbon (middle) and iron (bottom). The respective standard deviations are also shown in the panels.}
\end{figure}

\begin{figure*}
\gridline{\fig{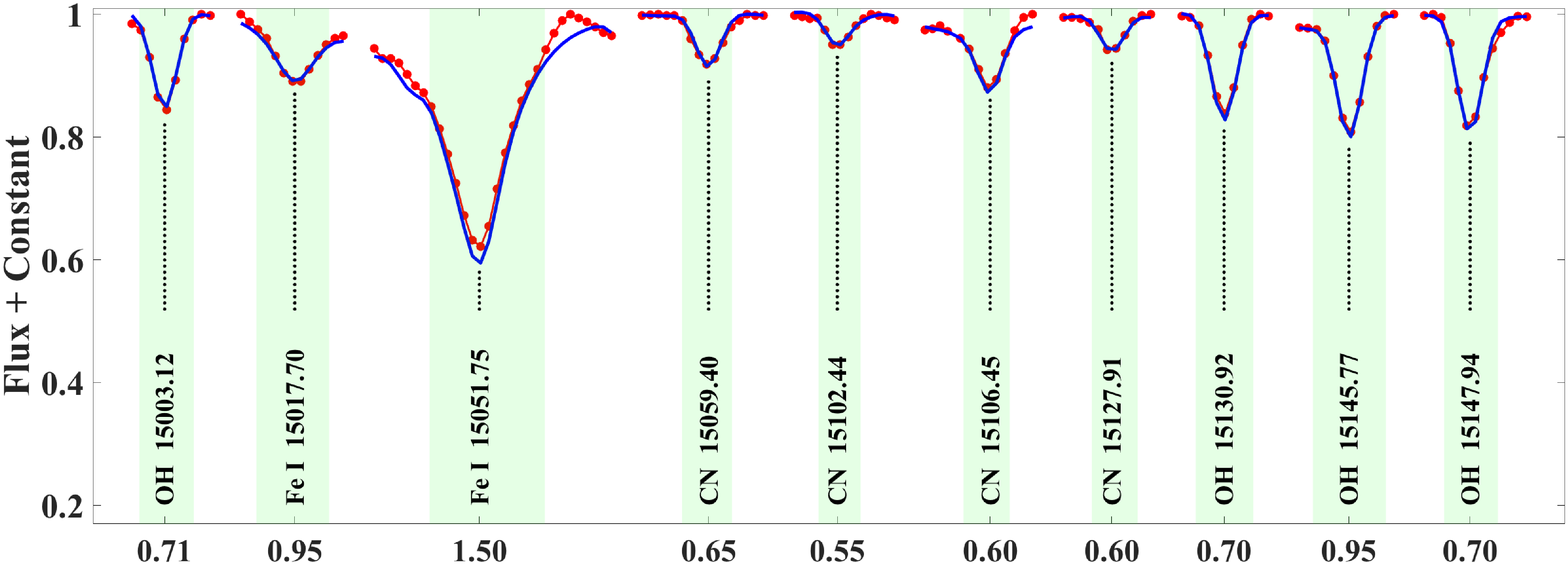}{0.95\textwidth}{}
}  

\vspace{-1.2cm}          
\gridline{\fig{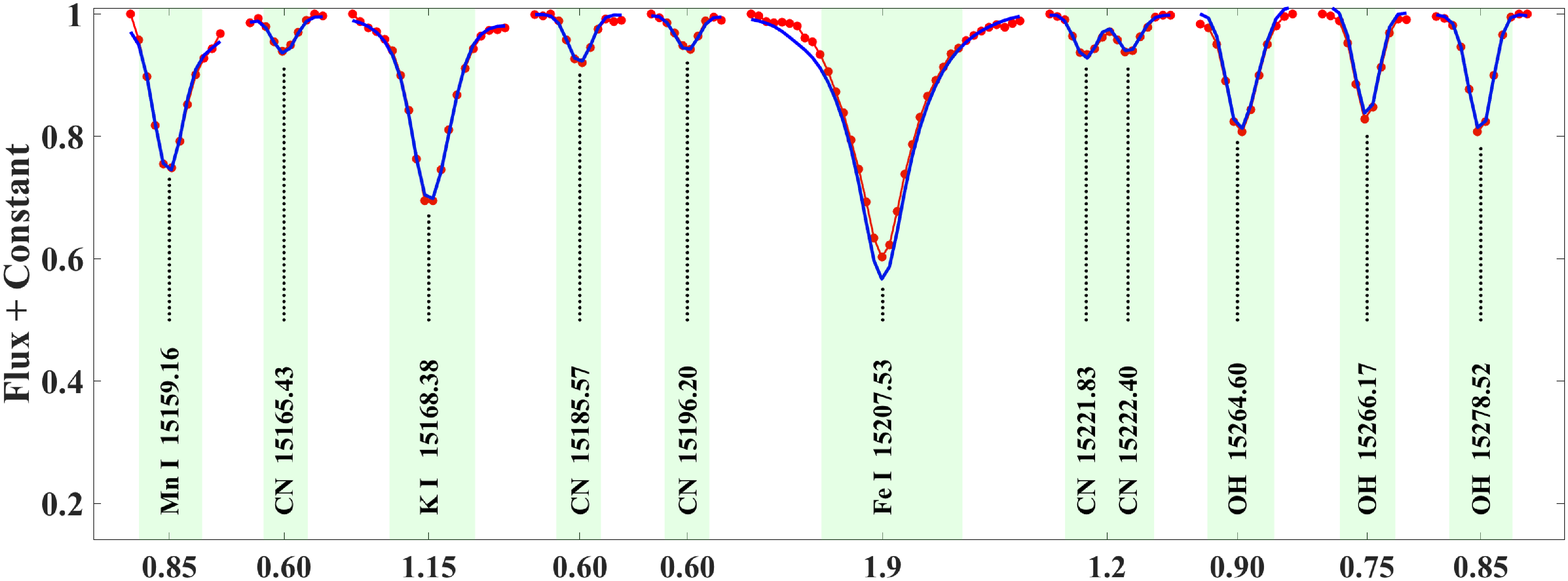}{0.95\textwidth}{}
}

\vspace{-1.2cm} 
\gridline{\fig{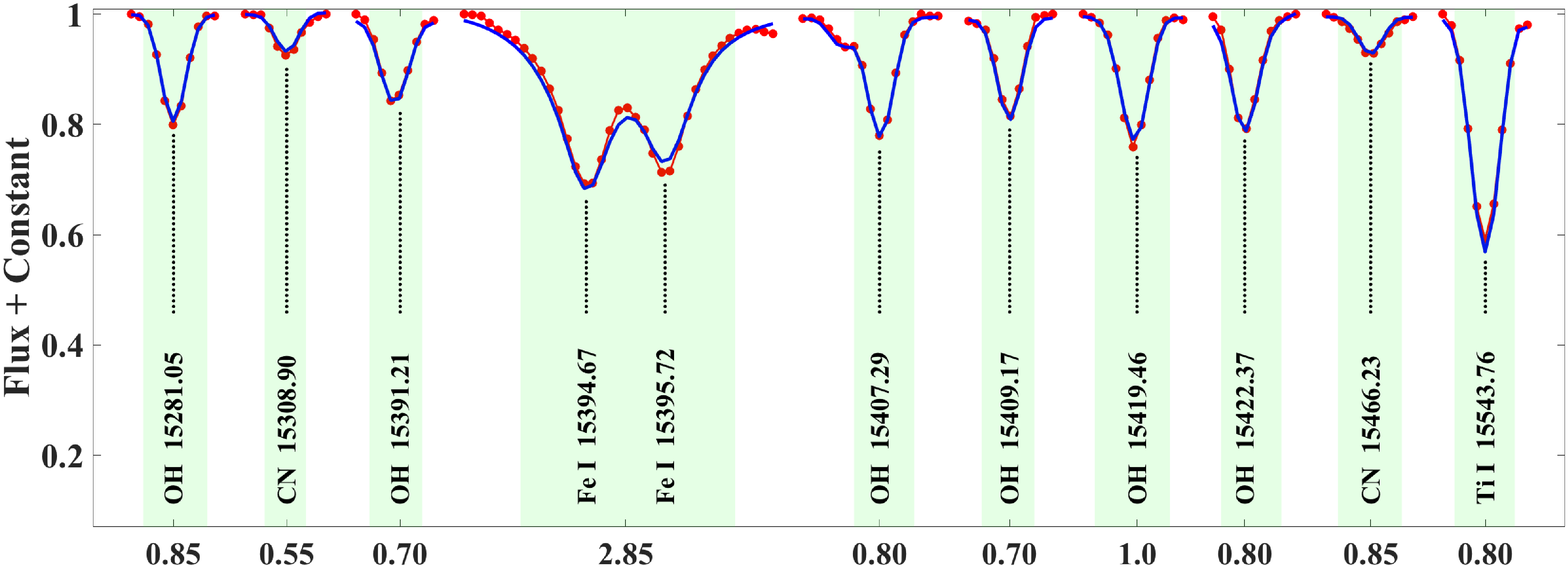}{0.95\textwidth}{}
}

\vspace{-1.2cm} 
\gridline{\fig{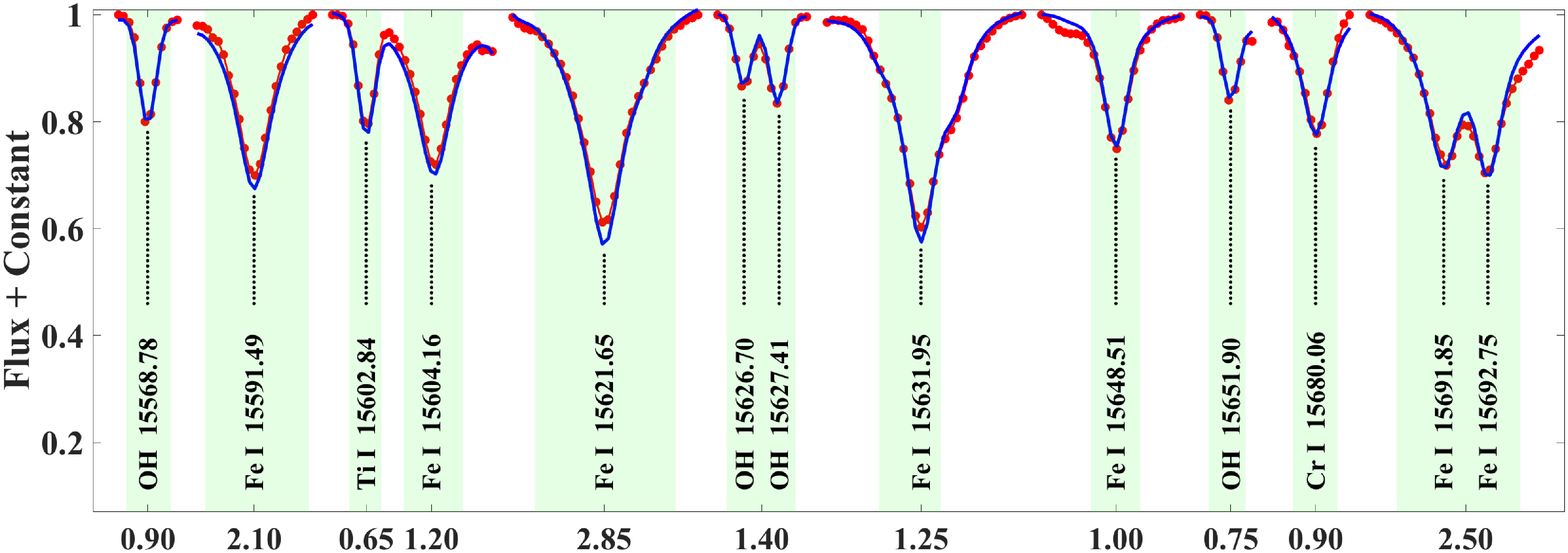}{0.95\textwidth}{}
}
\caption{Comparison between the renormalized, observed spectrum (red lines and dots) and the best-fit synthetic model (blue lines) over the selected spectral lines in the wavelength range between around 15003 {\AA}  and 15693 {\AA}. The shaded regions show the  $\chi$$^\textrm{\footnotesize{2}}$ windows used in the model-fit process. The numbers on the x-axes are the widths of the corresponding $\chi$$^\textrm{\footnotesize{2}}$ windows in angstroms.}
\end{figure*}

\begin{figure*}
\gridline{\fig{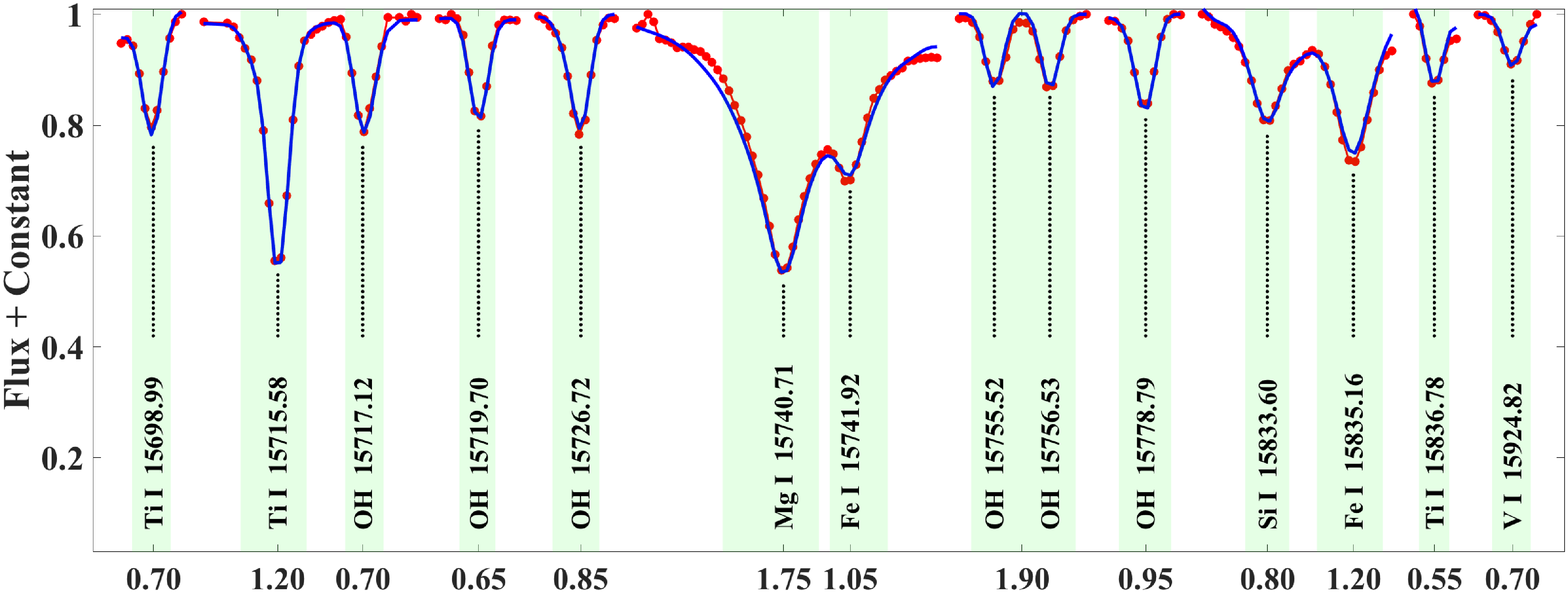}{0.95\textwidth}{}
}  

\vspace{-1.13cm}          
\gridline{\fig{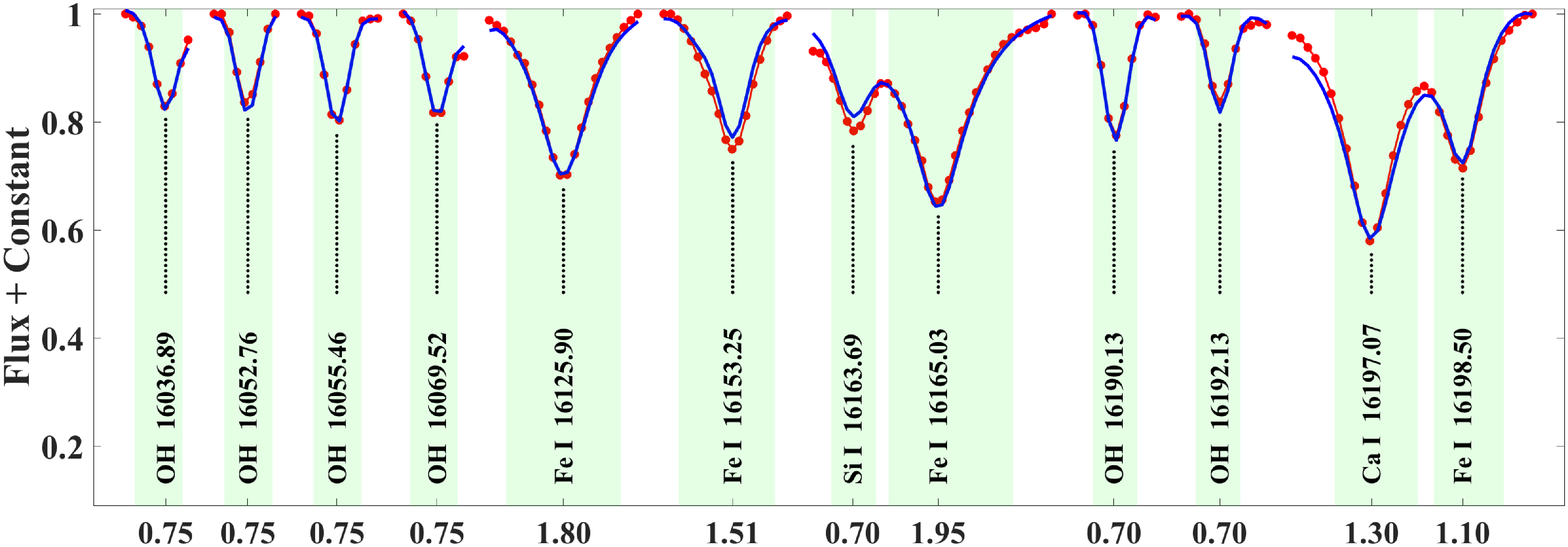}{0.95\textwidth}{}
}

\vspace{-1.13cm} 
\gridline{\fig{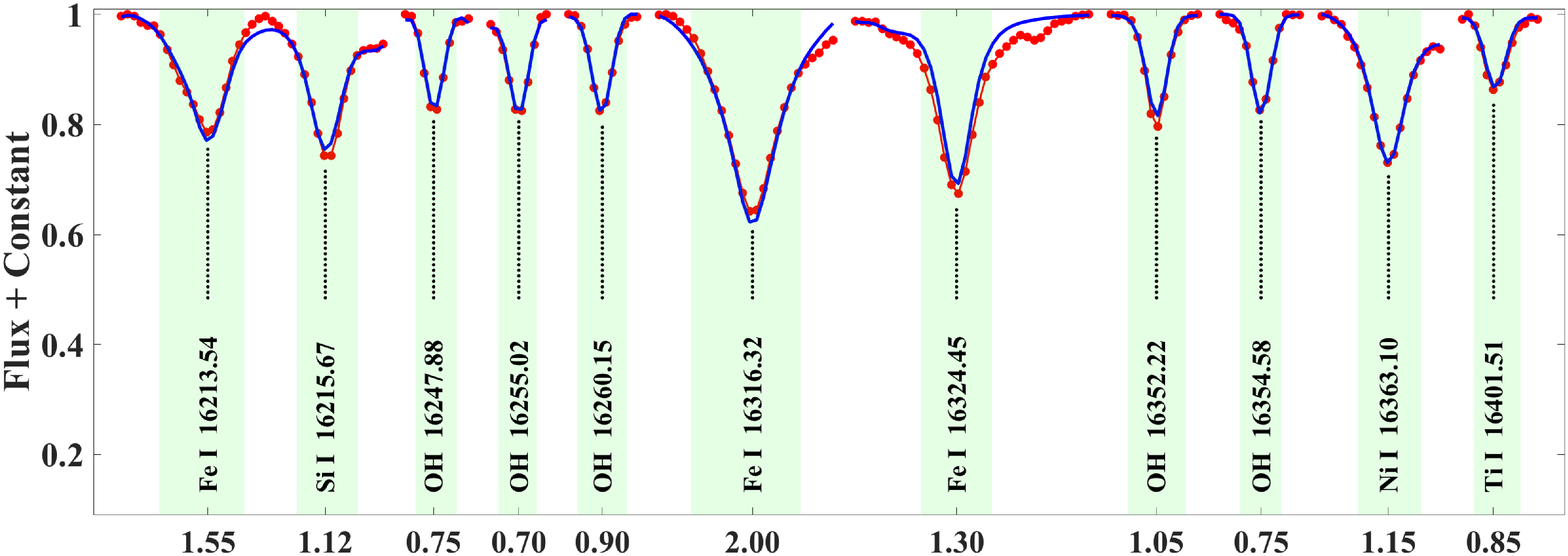}{0.95\textwidth}{}
}

\vspace{-1.13cm} 
\gridline{\fig{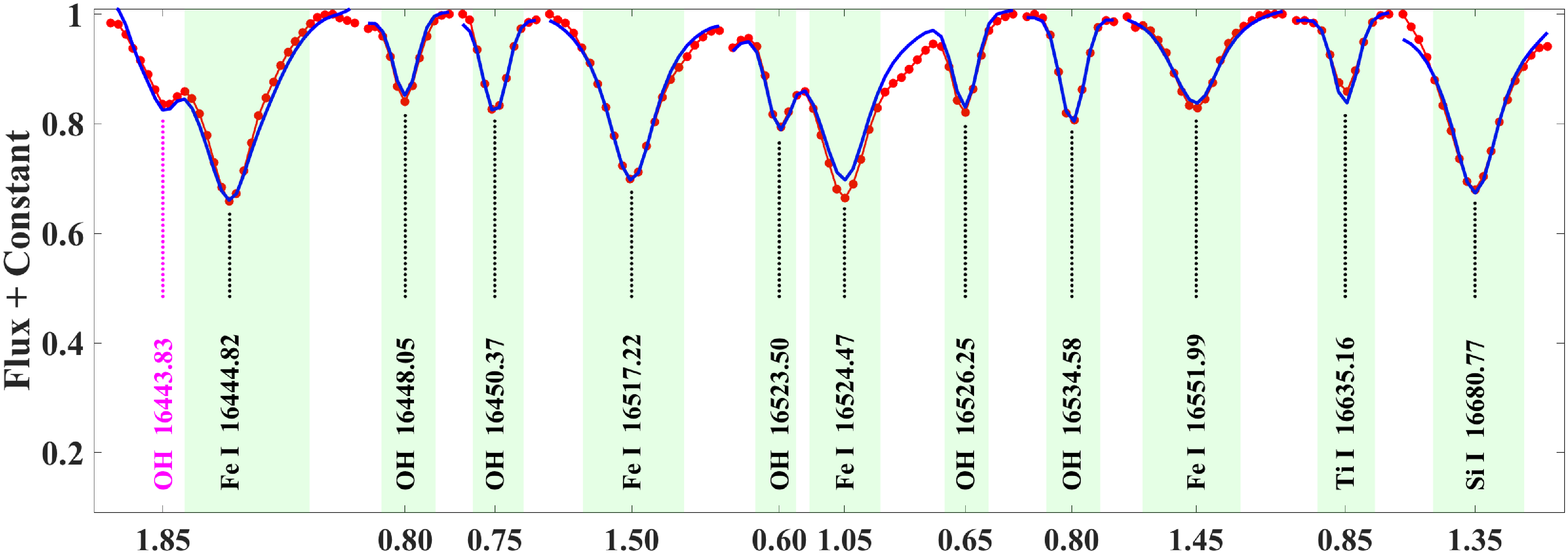}{0.95\textwidth}{}
}
\caption{Identical to Figure 4, except showing the spectral lines in the wavelength range between around 15698 {\AA}  and 16681 {\AA}. The line labeled in magenta was not used in the model-fit process.
}
\end{figure*}

\begin{figure*}
\gridline{\fig{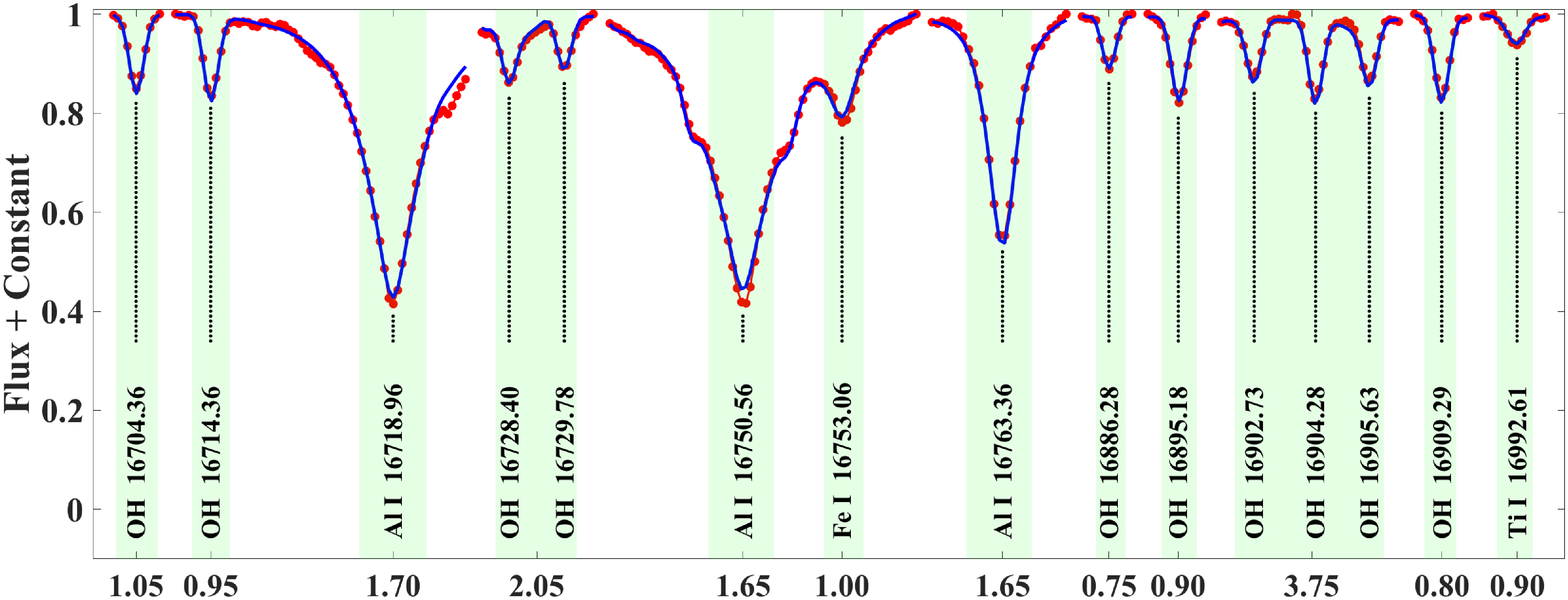}{0.95\textwidth}{}
}  

\vspace{-1.13cm}          
\gridline{\fig{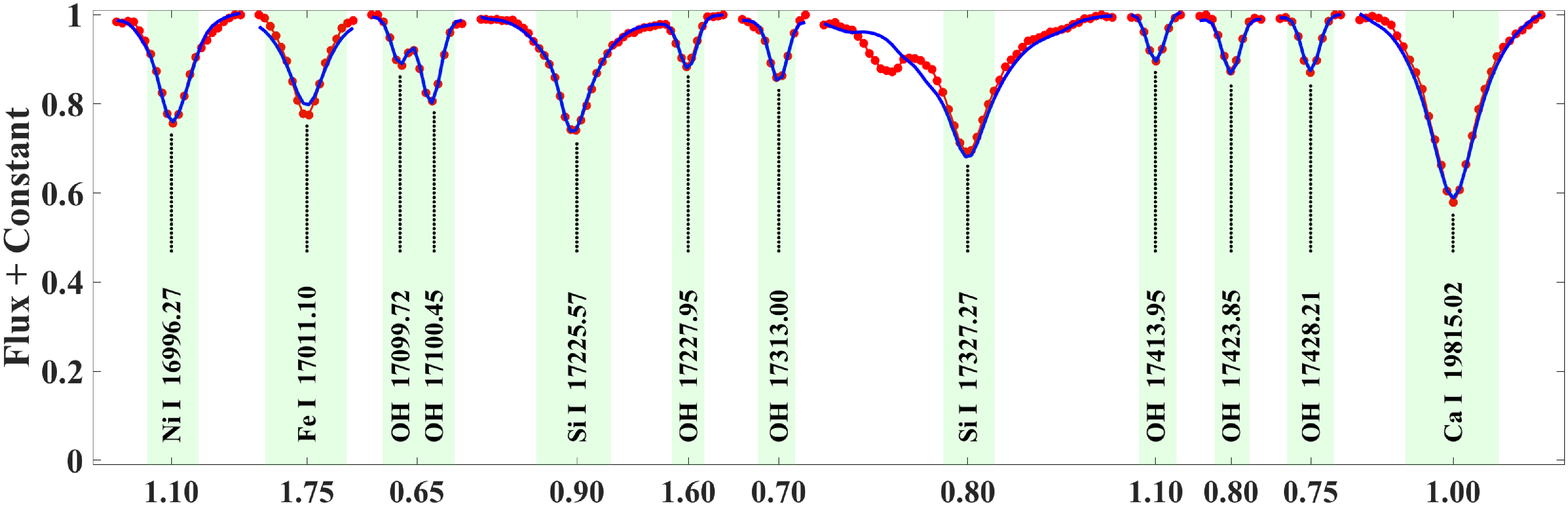}{0.95\textwidth}{}
}

\vspace{-1.13cm} 
\gridline{\fig{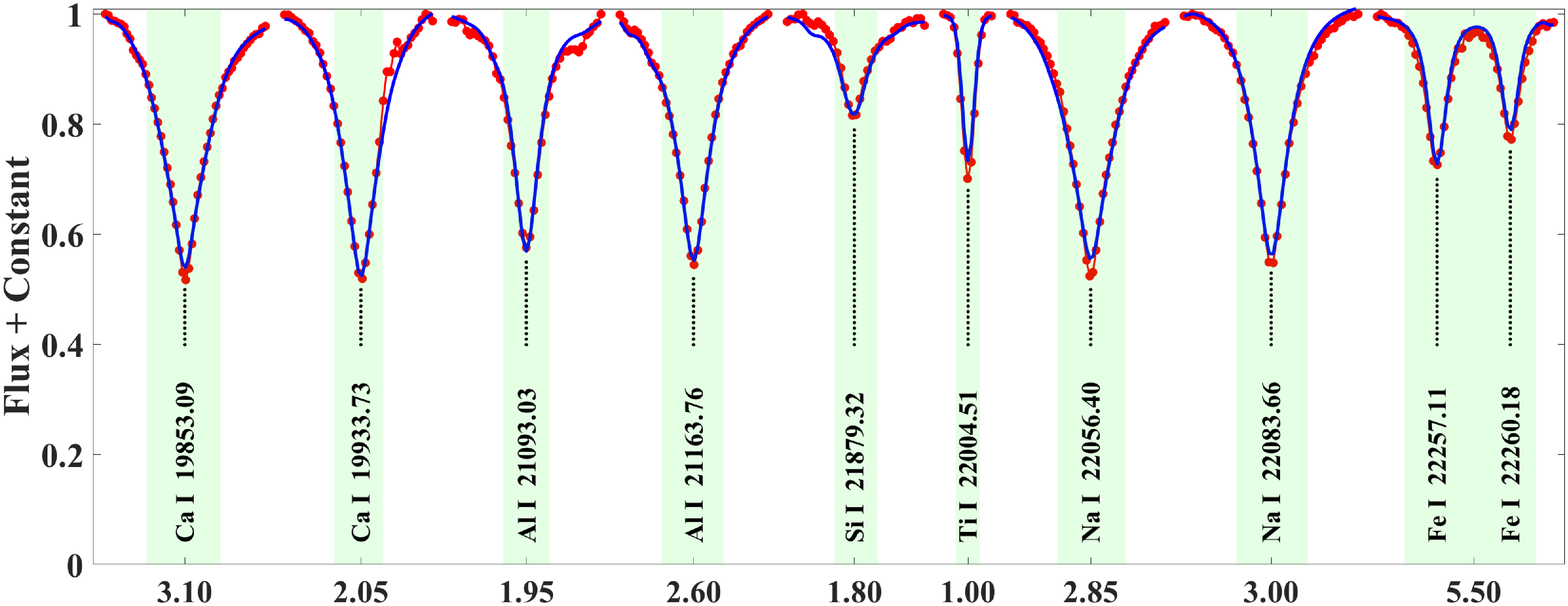}{0.95\textwidth}{}
}

\vspace{-1.13cm} 
\gridline{\fig{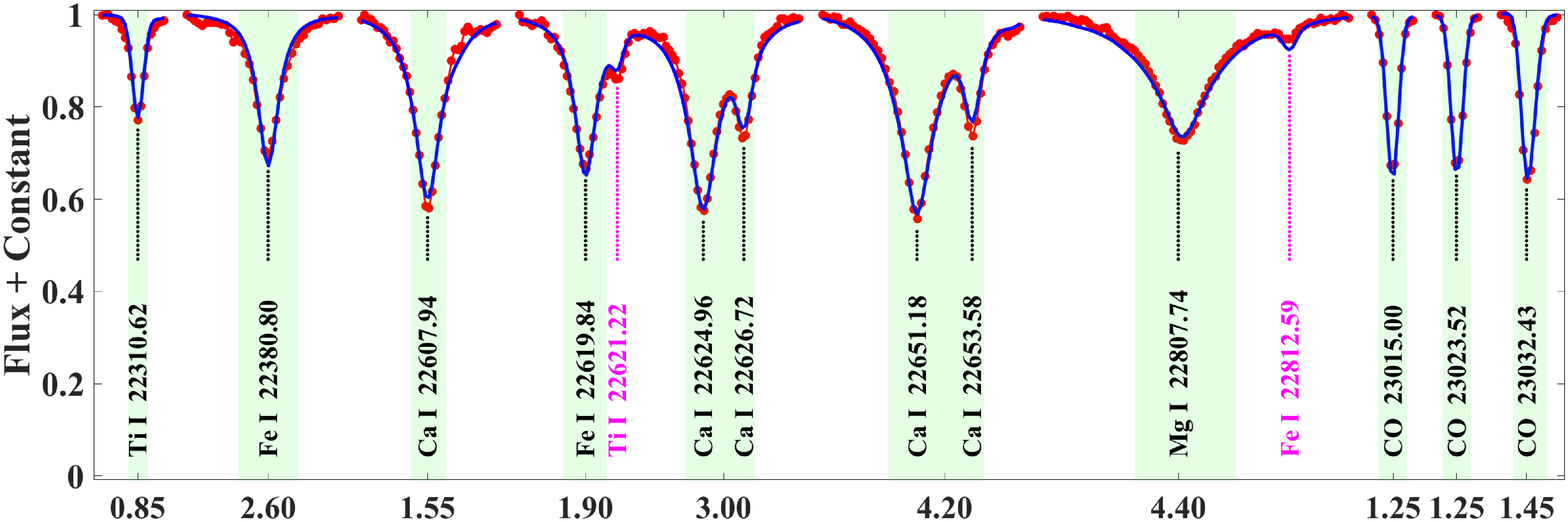}{0.95\textwidth}{}
}
\caption{Identical to Figure 4, except showing the spectral lines in the wavelength range between around 16704 {\AA}  and 23033 {\AA}. The lines labeled in magenta were not used in the model-fit process.}
\end{figure*}

\begin{figure*}
\gridline{\fig{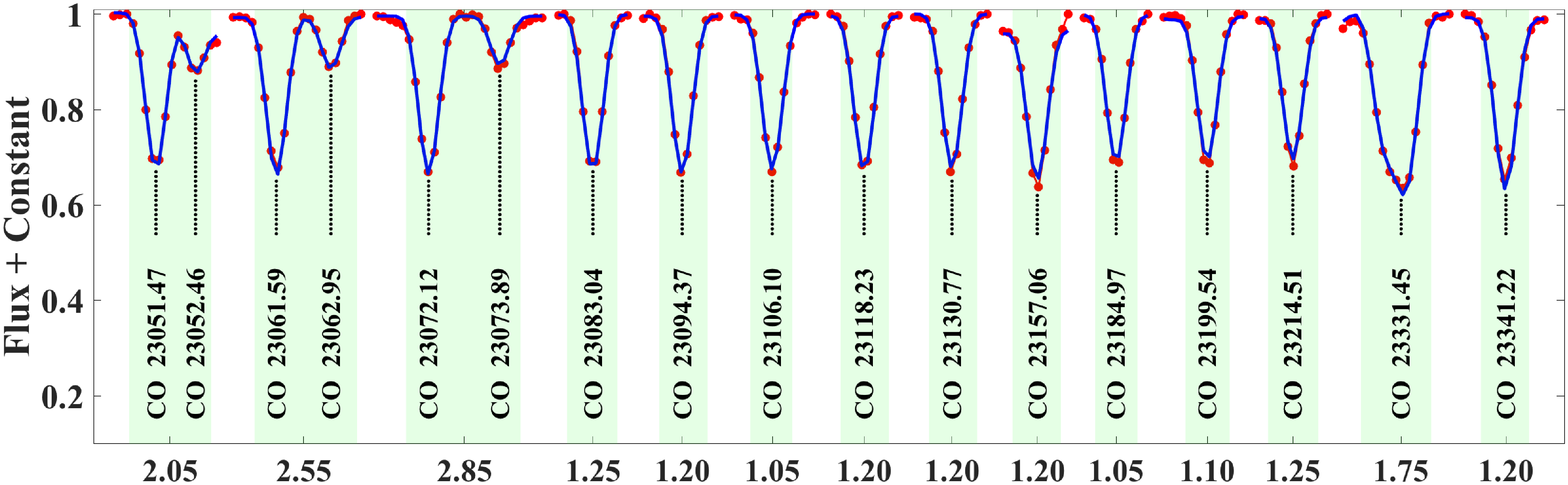}{0.95\textwidth}{}
}  

\vspace{-1.13cm}          
\gridline{\fig{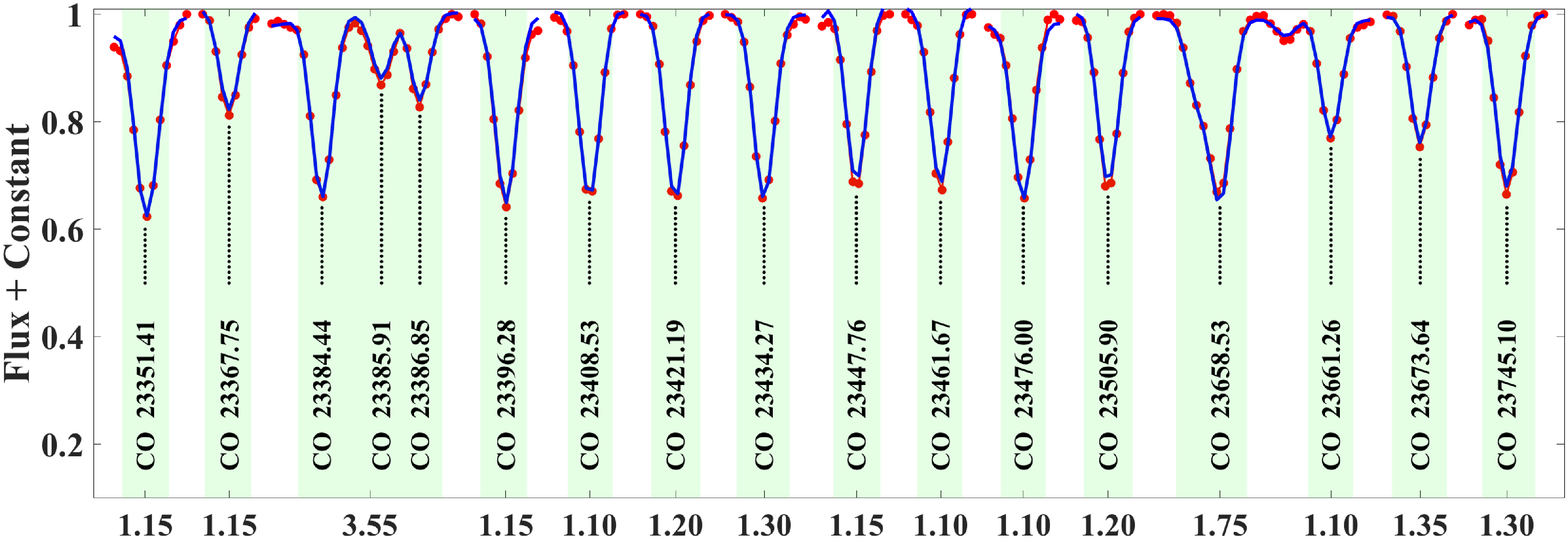}{0.95\textwidth}{}
}

\vspace{-1.13cm} 
\gridline{\fig{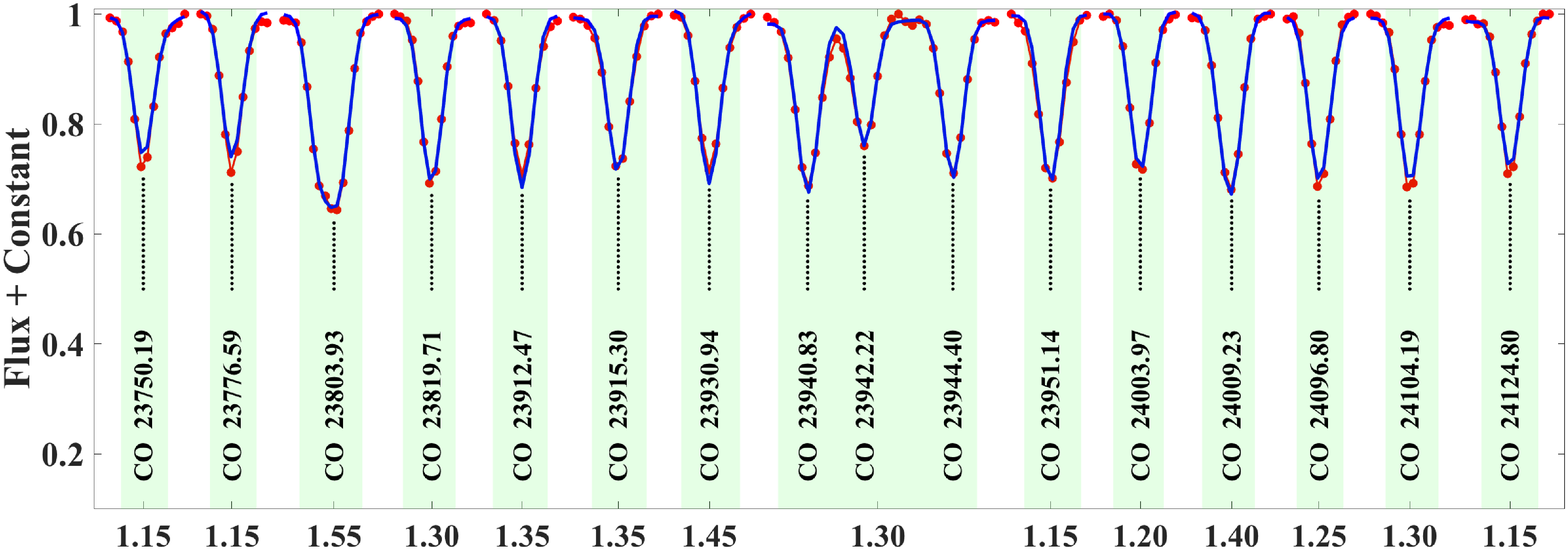}{0.95\textwidth}{}
}

\vspace{-1.13cm} 
\gridline{\fig{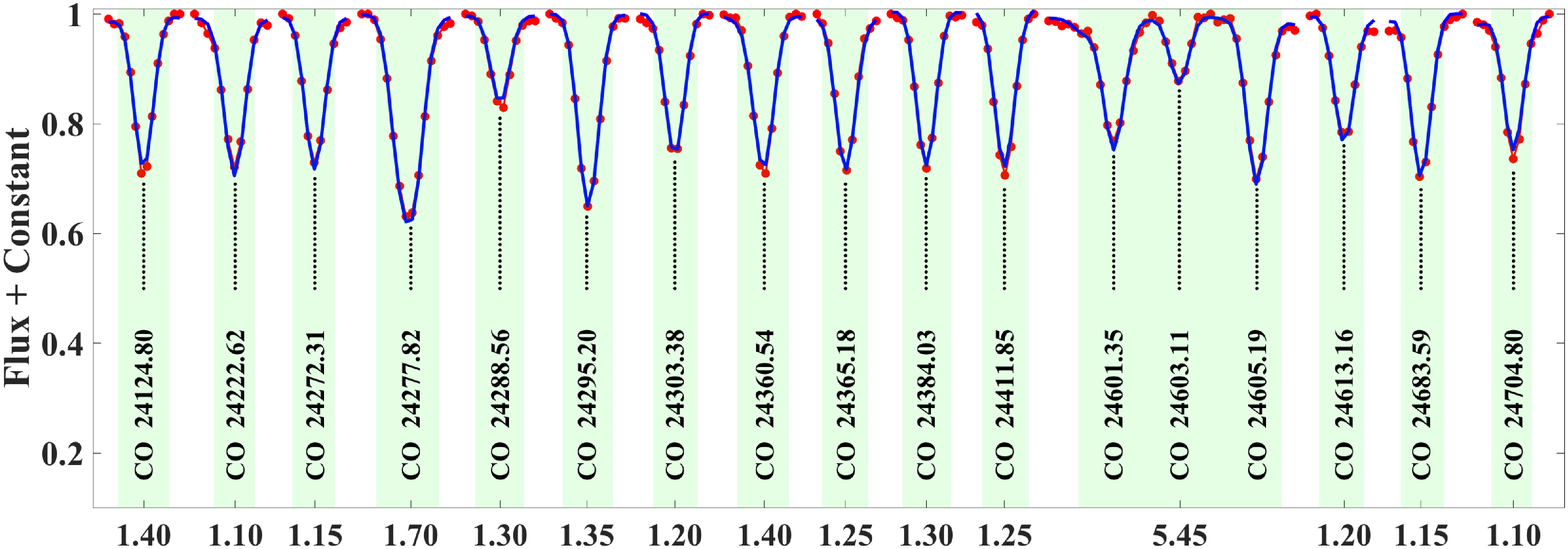}{0.95\textwidth}{}
}
\caption{Identical to Figure 4, except showing the spectral lines in the wavelength range between around 23051 {\AA}  and 24705 {\AA}.}
\end{figure*}

We found an excellent agreement between the observed spectrum and the best-fit model (i.e., the model associated with the physical parameters and the best-fit elemental abundances of the target) over the majority of the studied lines ($\simeq$80 $\%$). There is also a reasonable consistency between the observed spectrum and the best-fit synthetic model over the remaining lines used in the analysis. Figures 4-7 show 209 lines (out of 232 lines used in the fitting) that indicate a good match between the renormalized, observed flux (red lines and dots)  and the best-fit model (blue lines). The shaded regions show the $\chi$$^\textrm{\footnotesize{2}}$ windows used in the minimization routine. The numbers on the x-axes are the widths of the corresponding $\chi$$^\textrm{\footnotesize{2}}$ windows in angstroms. In these figures, the lines were sorted in order of increasing wavelength, and since the main purpose is to show the quality of the fit, the lines were shifted to the same flux level (i.e., unity, regardless of the significant flux depression over some lines) for better presentation. The lines that were passed through the model-fit routine are labeled in black while the lines that were not chosen for the fitting analysis are labeled in magenta.  The latter were excluded because, during the line selection, they did not seem to be strong enough, or were evidently inconsistent with the initial estimate of the best-fit model. It should  be recalled that the best-fit abundances used in synthesizing the best-fit model are the average of abundances inferred from multiple lines (if applicable). Given the scatter (and the resulting standard errors shown in Table 1) around the average abundances, this can cause  discrepancies  between the observed spectrum and  best-fit model over some spectral lines.

 \begin{deluxetable*}{lccccccc} [h!]
 \tablenum{1}
\tablecaption{The chemical abundances and their corresponding uncertainties for the fifteen studied elements}  
\tablewidth{0pt}
\tablehead{
{\footnotesize{Species}}   &  {\footnotesize{\textit{N}}}  &{\footnotesize{[X/H]}} &
{\footnotesize{A(X)}} & {{$\sigma$$_\textrm{ran}$}} & {$\sigma$$_\textrm{\footnotesize{Teff}}$}  & {$\sigma$ $_\textrm{\footnotesize{[M/H]}}$} & {$\sigma$$_\textrm{\footnotesize{tot}}$}}
\startdata
C (CO)	& 70	 & $-$0.066 &  8.324	& 0.001 & +0.044  & +0.029 & 0.053\\
N (CN)	& 12	 & +0.012	 & 7.792	& 0.005	& +0.024	& $-$0.006	 & 0.025\\
O (OH)	& 60	 & $-$0.037 & 8.623	& 0.000	& +0.085	& +0.005   &0.085\\
Na	& 2	 & +0.161	 & 6.331	& 0.047	& +0.067  &   $-$0.048	&  0.095\\
Mg	& 3	&  $-$0.124 & 7.406	& 0.037	& $-$0.029  &   +0.016	&  0.050\\
Al	& 5	& +0.028	 & 6.398	& 0.008	& +0.047  &  $-$0.036	&  0.060\\
Si	& 10	 & $-$0.136 & 7.374	& 0.009	& $-$0.049	&  +0.024	&  0.055\\
K	& 1	 & +0.014	 & 5.094	& ---  &  +0.031	&  +0.004	 & 0.031\\
Ca	& 10	 & $-$0.022	 & 6.288	& 0.007	& +0.048	&  $-$0.018	 & 0.052\\
Ti	& 10	&  +0.000	 & 4.900	& 0.007	& +0.065	&  +0.005	&  0.066\\
V	& 1	&  $-$0.061	&  3.939	& ---  &  +0.051	&  +0.028	 & 0.058\\
Cr	& 1	&  +0.238	 & 5.878	& --- &  +0.006	&  +0.019	 & 0.020\\
Mn	& 1	&  $-$0.226	 & 5.164	& --- &  $-$0.013	 & +0.013	 & 0.018\\
Fe	& 44	 & $-$0.071	 & 7.379	& 0.002    &  $-$0.023  & +0.014	&  0.027\\
Ni	& 2	 &  $-$0.140	&  6.090	& 0.004  & $-$0.028	& +0.019	&  0.034\\
\enddata
\tablecomments{$\sigma$$_\textrm{\footnotesize{ran}}$ shows the standard error of the mean (std/$\sqrt{N}$), and $\sigma$$_\textrm{\footnotesize{Teff}}$ and $\sigma$$_\textrm{\footnotesize{[M/H]}}$ indicate the systematic errors resulting from varying T$_\textrm{\footnotesize{eff}}$ and [M/H] by their corresponding uncertainties (70 K and 0.09 dex), respectively. Due to their unknown statistical errors, the total uncertainties of the four elements K, V, Cr, and Mn  are underestimated.
} 
\end{deluxetable*}

\subsection{Error Analysis}  
The standard (random) error of the mean ($\sigma$$_\textrm{\footnotesize{ran}}$), i.e., std/$\sqrt{N}$, where std is the standard deviation of the abundances derived from  \textit{N} lines of each particular element,  is  shown in the fifth column of Table 1. It is to be noted that the random error is not applicable for the elements with single-line inferred abundances, i.e., K, V, Cr, and Mn.
  
The systematic uncertainties of the derived abundances resulting from the errors in stellar parameters were estimated using the approach described in Souto et al. (2016, 2017). We changed the effective temperature and metallicity by the errors reported in Piaulet et al. (2021) one at a time, i.e.,  T$_\textrm{\footnotesize{eff}}$ + 70 K = 4495 K and  [M/H] + 0.09  dex = 0.11 dex,  and then obtained the elemental abundances for each case using our iterative model-fit procedure.  The error of surface gravity (0.012 dex) is too small to make a noticeable impact on the elemental abundances, and we did not include the effect of  this parameter in the systematic errors. We also found a negligible change in the inferred abundances due to the variation of the adopted value of microturbulence parameter by 10{\%} (i.e., $\Delta\xi$ = 0.125 km/s), and the effect of this parameter was excluded from the systematic error estimation as well.

The abundance systematic errors due to the perturbation of T$_\textrm{\footnotesize{eff}}$ ($\sigma$$_\textrm{\footnotesize{Teff}}$) and [M/H]  ($\sigma$$_\textrm{\footnotesize{[M/H]}}$) are presented in the sixth and seventh columns of Table 1, respectively.  The last column of Table 1 shows the quadrature sum of the random ($\sigma$$_\textrm{\footnotesize{ran}}$) and systematic ($\sigma$$_\textrm{\footnotesize{Teff}}$ and $\sigma$$_\textrm{\footnotesize{[M/H]}}$) errors as the total uncertainty ($\sigma$$_\textrm{\footnotesize{tot}}$) for each element. The inferred abundance versus atomic number for  the fifteen elements, along with the total abundance uncertainties shown as error bars, are also illustrated in Figure 8.  We stress that due to the unknown statistical errors associated with  K, V, Cr, and Mn, the total uncertainties of these four elements are underestimated. In particular, the abundance of  V inferred from only one relatively weak line may be uncertain considerably beyond its reported total uncertainty. Nevertheless, the overall  values of $\sigma$$_\textrm{\footnotesize{tot}}$ indicate the high quality of our abundance measurement technique that  results in the detailed chemical abundances of the most essential elements with precision $<$ 0.1 dex for a planet-host star. Our analysis means that WASP-107 is now the coolest host star in the JWST Cycle 1 exoplanet sample with precise measured chemical abundances.

\begin{figure*}
\gridline{\fig{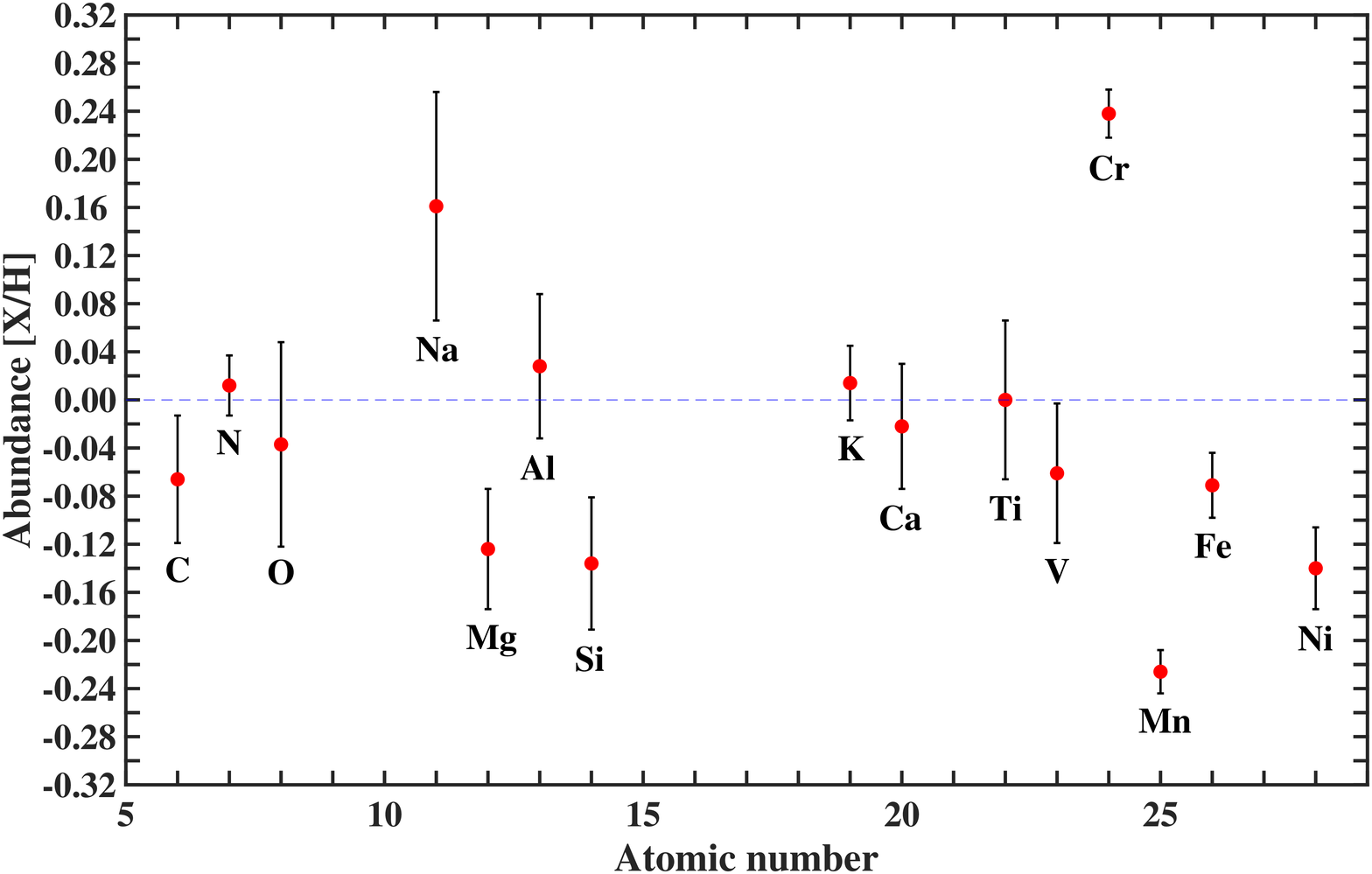}{0.8\textwidth}{}
}  
\caption{Abundance versus atomic number for the fifteen analyzed elements (that are labeled in the plot). Error bars show the total uncertainties of the inferred abundances.}
\end{figure*}

\section{Discussion}  
Elemental abundances of planet-host stars provide critical insight into the formation and properties of the  orbiting planets. In particular, certain abundance ratios of host stars such as C/O and Mg/Si serve as proxies for the  formation region, structure, and mineralogy of the planets. The elemental abundances of our target star are near-solar values, which are not surprising given the near-solar metallicity ([M/H] = +0.02 dex) of the star.  However,  the other JWST target stars are not assumed to have near-solar abundances, so it is critical to measure the composition of all stars in the sample once the atmospheric abundances of the respective planets become available. In addition, we  found  C/O = 0.50 $\pm$ 0.11  and  Mg/Si = 1.08 $\pm$ 0.18 ratios for the target, which are consistent with the solar values (for reference, the solar ratios are (C/O)$_\textrm{\footnotesize{$\odot$}}$ = 0.54 $\pm$ 0.09 and (Mg/Si)$_\textrm{\footnotesize{$\odot$}}$ = 1.05 $\pm$ 0.24; Grevesse et al. 2007). We note the uncertainties of individual elemental abundances and abundance ratios inferred from this study are quite comparable to those reported for 17 hotter (FGK) dwarfs in the exoplanet-focused Cycle 1 JWST observer programs  using equivalent width analysis (Kolecki \& Wang 2022).  The uncertainties of our derived chemical abundances are also comparable to those reported for 1111 FGK stars from the HARPS GTO planet search program again using  equivalent width analysis (Adibekyan et al. 2012b; Delgado Mena et al. 2017 and 2021; Costa Silva et al. 2020). The accuracy of our analysis is mostly limited by  the errors of the star's physical parameters, which give rise to uncertainties in the inferred elemental abundances, both statistical  and systematic. Improvements upon stellar parameter determination would definitely decrease the uncertainties of our chemical abundance measurements.

It is believed that gas giant planets that accrete gas beyond the H$_\textrm{\footnotesize{2}}$O  ice line have higher C/O ratios, as compared to the values of their parent stars. This suggests that beyond the water ice line, most oxygen atoms are trapped in solid water ice particles, leaving a large fraction of carbon in the form of gas. In contrast, giant planets that accrete significant amounts of solid planetesimals within the H$_\textrm{\footnotesize{2}}$O  ice line indicate lower C/O ratios with respect to those of the host stars ({\"O}berg et al. 2011). This is consistent with more recent studies (Espinoza et al. 2017; Lothringer et al. 2021) which show an inverse correlation between C/O ratios and heavy-element enrichment of giant planets.  Based on the HST spectroscopic analysis of super-Neptune WASP-107b,  the planet  seems to have a low C/O ratio, which may be due to the accretion of water-rich planetesimals (Espinoza et al. 2017; Mordasini et al. 2016; Kreidberg et al. 2018). Since the planet is inside the H$_\textrm{\footnotesize{2}}$O  ice line, a lower planetary C/O is expected. The JWST observations of this planet will, however, offer a more accurate measurement of metal content and C/O ratio, which  allows a proper comparison between the stellar and planetary chemical properties, and subsequently an estimate of the planet original location. 

Although the C/O ratio varies over different parts of the interstellar medium (ISM), stars with high C/O values (C/O $>$ 0.8) appear to be rare. The paucity of carbon-rich stars has been  confirmed with different stellar samples in the Solar neighborhood (Brewer \& Fischer 2016; Su\'{a}rez-Andr\'{e}s et al. 2018).
Using $\sim$ 850 nearby F, G, and K dwarfs (with 156 known planet hosts), Brewer \& Fischer (2016) found a  median of 0.47 for the C/O values. The Mg/Si ratios showed a broader distribution that peaked around the  median, i.e., Mg/Si =  1.02 (near the solar value),  with  about 60{\%} of stars having 1 $\leqslant$ Mg/Si $<$ 2 and 40{\%} having Mg/Si $<$ 1. The mineralogical ratios C/O and Mg/Si  were later studied in greater detail with a sample of 99 solar-like plant hosts (Su\'{a}rez-Andr\'{e}s et al. 2018). All stars showed  C/O $<$ 0.8, and the distribution peaked at $\sim$ 0.47, and only $\sim$ 15{\%} of stars had C/O $<$ 0.4. The sample was then divided into two groups; host stars with low-mass planets ($\leqslant$ 30 \textit{M$_\textrm{\footnotesize{$\oplus$}}$}) and  host stars with high-mass planets ($>$ 30 \textit{M$_\textrm{\footnotesize{$\oplus$}}$}), which had an average C/O ratio  of  0.46 and 0.50, respectively. Among stars with high-mass companions, 86{\%}  had ratios in the range 0.4 $<$ C/O $<$ 0.8, and the remaining 14{\%} had C/O $<$ 0.4. Nearly the same fractions were found for the C/O ratios of stars hosting low-mass planets. In regard to Mg/Si ratios, 85{\%} of host stars with high-mass planets showed 1.0 $<$ Mg/Si $<$ 2.0, while the rest of the subsample presented Mg/Si $<$ 1.0. All stars with low-mass companions had 1.0 $<$ Mg/Si $<$ 2.0. 

Some other works have also found a diversity in C/O and Mg/Si ratios of star samples, which suggest various types of planetary systems with different formation pathways. Nevertheless, our target star WASP-107 has a C/O ratio (0.50) close to  the average values of samples analyzed in Brewer \& Fischer (2016) and Su\'{a}rez-Andr\'{e}s et al. (2018), falling within the C/O distribution of  the majority of stars in both samples.  The Mg/Si ratio of our star (1.08) is also comparable with the values of majority of stars in the two studied samples. In particular, the mass of WASP-107b is  slightly higher than the border mass (30 \textit{M$_\textrm{\footnotesize{$\oplus$}}$}) between the two stellar groups in Su\'{a}rez-Andr\'{e}s et al. (2018), which puts our target into the group with high-mass planets. The target's Mg/Si ratio is in the range of 1.0 $<$ Mg/Si $<$ 2.0 valid for the 85{\%} of  stars in this subset of high-mass companions.  

In general, the distribution of Si among carbide and oxide species is controlled by C/O ratio (e.g., Bond et al. 2010; also see Brewer et al. 2016 and references therein). If  C/O $>$ 0.8, Si exists in solid form as SiC. In addition, graphite and TiC are also formed. If C/O $<$ 0.8, Si is present in rock-forming minerals such as SiO$_\textrm{\footnotesize{4}}^\textrm{\footnotesize{4$-$}}$ or SiO$_\textrm{\footnotesize{2}}$, which serve as seeds for Mg silicates whose compositions are specified by Mg/Si ratio. In particular, for Mg/Si $<$ 1.0, Mg forms orthopyroxene (MgSiO$_\textrm{\footnotesize{3}}$) while the remaining Si  is present in other silicates such as feldspars (e.g., CaAl$_\textrm{\footnotesize{2}}$Si$_\textrm{\footnotesize{2}}$O$_\textrm{\footnotesize{8}}$ and NaAlSiO$_\textrm{\footnotesize{8}}$) or olivine (Mg$_\textrm{\footnotesize{2}}$SiO$_\textrm{\footnotesize{4}}$). For 1.0 $<$ Mg/Si $<$ 2.0, Mg is equally distributed between olivine and pyroxene. Given the C/O and Mg/Si ratios of our star target, an equal proportion of olivine and pyroxene is expected for the rocky core of the planet WASP-107b.

It is worth mentioning that there are some other key elements such as sulfur which has also proved to be crucial in determining  the composition and chemistry of gas giant exoplanets (Tsai et al. 2022). However, we could not detect the atomic S lines in the NIR spectra of our target, as these lines are also weak and blended with other atomic and molecular lines, though these lines can be  measurable over some wavelength ranges (e.g., near 6743 {\AA}, 6748 {\AA}, and 6757 {\AA}) in the  high-resolution,  optical spectra of low-temperature stars (e.g., Perdigon et a. 2021). 

In summary, a critical  approach to study exoplanet properties is the scrutiny of  the parent stars. The abundance analysis of the K dwarf WASP-107 presented here is a pilot study that opens the way to detailed abundance measurements of all other  JWST's cooler exoplanet-host stars. The chemical abundances of these host stars can then  be compared to those of the respective planets from the forthcoming JWST spectroscopic analyses, which could reveal pivotal clues on the formation, evolution, and characterization of exoplanets. It is important to note that MARCS model atmospheres have been shown to be capable of sufficiently modeling the spectra of cool stars down to T$_\textrm{\footnotesize{eff}}$ $\simeq$ 3200 K (Souto et al. 2018). As our follow-up study to test the accuracy of  chemical abundance measurements of  stars with T$_\textrm{\footnotesize{eff}}$ $<$ 3200 K  using MARCS model atmospheres,  we will use wide binary systems that consist of  mid-to-late type M dwarfs with more massive FGK-type companions. The comparison between the inferred abundances of the M dwarfs using MARCS models and those of the companions using other methods can allow us to evaluate the sufficiency  of these models in the spectroscopic analysis of very low-mass stars. It should be noted that M+FGK dwarf Binary systems have already been used to verify the abundance measurements of M dwarfs (Ishikawa et al. 2020; Souto et al. 2022).\\

We greatly thank the anonymous referee for the insightful comments and suggestions that improved the manuscript.
We are grateful for assistance provided by Thomas Masseron while the work was being undertaken. We also wish to thank Justin Cantrell and Jeremy Simmons for their technical support with the  high-performance computing system of the physics and astronomy department, Georgia State University, which was used for this study. NH and IJMC acknowledge support from NSF AAG grant 2108686 and from NASA ICAR grant NNH19ZDA001N. TN acknowledges  support from the Australian Research Council Centre of Excellence for All Sky Astrophysics in 3 Dimensions (ASTRO 3D), through project No. CE170100013. DS thanks the National Council for Scientific and Technological Development – CNPq. EM acknowledges financial support from Gobierno de Canarias and the European Regional Development Fund (ERDF) through project ProID2021010128.

\section{References}

\noindent 
\footnotesize{Abia, C., Tabernero, H. M., Korotin, S. A., et al. 2020, A\&A,  642, A227}

\noindent 
\footnotesize{Adibekyan, V. Zh., Santos, N. C., Sousa, S. G., et al. 2012a, A\&A, 543, A89}

\noindent 
\footnotesize{Adibekyan, V. Zh., Sousa, S. G., Santos, N. C., et al. 2012a, A\&A, 545, A32}

\noindent 
\footnotesize{Allart, R., Bourrier, V., Lovis, C., et al. 2019, A\&A, 623, A58}

\noindent 
\footnotesize{Alvarez, R. \& Plez, B. 1998, A\&A, 330, 1109}

\noindent 
\footnotesize{An, D., Pinsonneault, M. H.,  Masseron, T., et al. 2009, ApJ, 700, 523}

\noindent 
\footnotesize{Anderson, D. R., Collier Cameron, A., Delrez, L., et al. 2017, A\&A, 604, A110}

\noindent 
\footnotesize{Baily, E., \& Batygin, K. 2018, ApJL, 866, L2}

\noindent 
\footnotesize{Barber, R. J., Tennyson, J., Harris, G. J., \& Tolchenov, R. N. 2006, MNRAS, 368, 1087}

\noindent 
\footnotesize{Beaug{\'e}, C. \& Nesvorn{\'y}, D. 2013, ApJ, 763, 12}

\noindent 
\footnotesize{Becker, A. C., Agol, E., Silvestri, N. M., et al. 2008, MNRAS, 386, 412}

\noindent 
\footnotesize{Bensby, T.,  Gould, A.,  Asplund, M., et al. 2021, A\&A, 655, 117}

\noindent 
\footnotesize{Bond, J. C., O'Brien, D. P., \& Lauretta, D. S. 2010, ApJ, 715, 1050}

\noindent 
\footnotesize{Bonsor, A.,  Jofr\'{e}, P., Shorttle, O., et al. 2021, MNRAS, 503, 1877}

\noindent 
\footnotesize{Brewer, J. M., \& Fischer, D. A. 2016, ApJ, 831, 20}

\noindent 
\footnotesize{Brewer, J. M., Fischer, D. A., Valenti, J. A., \& Piskunov, N. 2016, ApJS, 225, 32}

\noindent 
\footnotesize{Brewer, J. M., \& Fischer, D. A. 2017, ApJ, 840, 121}

\noindent 
\footnotesize{Brooke, J. S. A., Ram, R. S., Western, C. M., et al. 2014, ApJS, 210, 23}

\noindent 
\footnotesize{Brugamyer, E., Dodson-Robinson, S. E., Cochran, W. D., \& Sneden, C. 2011, ApJ, 738, 97}

\noindent 
\footnotesize{Buchhave, L. A., Bizzarro, M., Latham, D. W., et al. 2014, Natur, 509, 593}

\noindent 
\footnotesize{Buchhave, L. A., \& Latham, D. W. 2015, ApJ, 808, 187}

\noindent 
\footnotesize{Costa Silva, A. R., Delgado Mena, E., \& Tsantaki, M. 2020, A\&A,  634, A136}

\noindent 
\footnotesize{Cristofari, P. I., Donati, J.-F., Masseron, T., et al. 2022, MNRAS, 511, 1893}

\noindent 
\footnotesize{Cushing, M. C., Vacca, W. D., \&  Rayner, J. T. 2004, PASP, 116, 362 }

\noindent 
\footnotesize{Dai, F., \& Winn, J. N. 2017, AJ, 153, 205}

\noindent 
\footnotesize{Davies, B., Kudritzki, R-P., \& Figer, D. F. 2010, MNRAS, 407, 1203} 

\noindent 
\footnotesize{Delgado Mena, E., Tsantaki, M., Adibekyan,  V. Zh., et al. 2017, 606, A94}

\noindent 
\footnotesize{Delgado Mena, E., Adibekyan, V. Zh.,  Santos, N. C., et al. 2021, A\&A,  655, A99}

\noindent 
\footnotesize{Dorn, C., Khan, A., Heng, K., et al. 2015, A\&A, 577, A83}

\noindent 
\footnotesize{Dressing, C. D., Hardegree-Ullman, K., Schlieder, J. E., et al. 2019, AJ, 158, 87}

\noindent 
\footnotesize{Dulick, M, Bauschlicher, Jr. C. W., Burrows, A., et al. 2003, ApJ, 594, 651}

\noindent 
\footnotesize{Espinoza, N., Fortney, J. J., Miguel, Y., Thorngren, D., \& Murray-Clay, R. 2017, ApJL, 838, L9}

\noindent 
\footnotesize{Fischer, D. A., \& Valenti, J. 2005, ApJ, 622, 1102}

\noindent 
\footnotesize{Fortney, J. J. 2012, ApJL, 747, L27}

\noindent 
\footnotesize{Gaia Collaboration, Brown, A. G. A., Vallenari, A., et al. 2018, A\&A, 616, A1}

\noindent 
\footnotesize{Gaia Collaboration, Brown, A. G. A., Vallenari, A., et al. 2021, A\&A, 649, A1}

\noindent 
\footnotesize{Goldman, A., 1982, ApOpt, 21, 2100}

\noindent 
\footnotesize{Gonzalez, G. 1997, MNRAS, 285, 403}

\noindent 
\footnotesize{Goorvitch, D., 1994, ApJS, 95, 535}

\noindent 
\footnotesize{Greene, T. P., Line, M. R., Montero, C., et al. 2016, ApJ, 817, 17}

\noindent 
\footnotesize{Grevesse, N., Asplund, M., \& Sauval, A. J. 2007, Space Sci. Rev., 130, 105}

\noindent 
\footnotesize{Heiter, U., Barklem, P., Fossati, L., et al. 2008, J. Phys. Conf. Ser., 130, 012011}

\noindent 
\footnotesize{Heiter, U., \& Luck, R. E. 2003, AJ, 126, 2015}

\noindent 
\footnotesize{Ishikawa, H. T.,  Aoki, W.,  Kotani, T., et al. 2020, PASJ, 72, 102 }

\noindent 
\footnotesize{Ishikawa, H. T., Aoki, W., Hirano, T. et al. 2022, AJ, 163, 72}

\noindent 
\footnotesize{Jofr\'e, P., Heiter, U., Soubiran, C., et al. 2015, A\&A, 582, A81}

\noindent 
\footnotesize{Johnson, J. A., Aller, K. M., Howard, A. W., \& Crepp, J. R. 2010, PASP, 122, 905}

\noindent 
\footnotesize{Kirk, J., Alam, M. K., L\'{o}pez-Morales, M., \& Zeng, L. 2020, AJ,  159, 115}

\noindent 
\footnotesize{Kolecki, J. R., \& Wang, J. 2022, AJ, 164, 87}

\noindent 
\footnotesize{Kreidberg, L., Line, M. R., Thorngren, D., et al. 2018, ApJL, 858, L6}

\noindent
\footnotesize{Kupka, F. G., Ryabchikova, T. A., Piskunov, N. E., Stempels, H. C., \& Weiss,
W. W. 2000, Balt. Astron., 9, 590}

\noindent
\footnotesize{Kurucz, R. L. 1995, ASPCS, 78, 205}

\noindent
\footnotesize{Larimer, J. W. 1975, Geochimica et Cosmochimica Acta, 39, 389}

\noindent
\footnotesize{Lecavelier Des Etangs, A. 2007, A\&A, 461, 1185}

\noindent
\footnotesize{Lee, J.-J., Gullikson, K., \& Kaplan, K. 2017, Igrins/Plp 2.2.0, Zenodo}

\noindent
\footnotesize{Lindgren, S., Heiter, U., \& Seifahrt, A. 2016, A\&A, 586, A100}

\noindent
\footnotesize{Liu, F., Yong, D., Asplund, M., et al.  2018, A\&A, 614, A138}

\noindent
\footnotesize{Lothringer, J. D., Rustamkulov, Z., Sing, D. K., et al. 2021, ApJ, 914, 12}

\noindent
\footnotesize{Marfil, E., Tabernero, H. M., Montes, D., et al. 2021, A\&A, 656, A162}

\noindent
\footnotesize{Mayor, M., \& Queloz, D. 1995, Nature, 378, 355}

\noindent
\footnotesize{Mordasini, C., Alibert, Y., Benz, W., \& Naef, D. 2009, A\&A, 501, 1161}

\noindent
\footnotesize{Mordasini, C., van Boekel, R., Molli\`{e}re, P., Henning, T., \& Benneke, B. 2016, ApJ, 832, 41}

\noindent
\footnotesize{Mortier, A., Santos, N. C., Sousa, S. G., et al. 2013, A\&A, 557, A70}

\noindent
\footnotesize{Muirhead, P. S., Veyette, M. J., Newton, E. R., et al. 2020, AJ, 159, 52}

\noindent
\footnotesize{Mulders, G. D., Pascucci, I., Apai, D., Frasca, A., \& Molenda-{\.Z}akowicz, J. 2016, AJ, 152, 187}

\noindent
\footnotesize{Nagar, T., Spina, L., \& Karakas, A. I. 2020, ApJL, 888, L9}

\noindent
\footnotesize{\"Oberg, K. I., Murray-Clay, R., \& Bergin, E. A. 2011, ApJL, 743, L16}

\noindent
\footnotesize{Oh, S., Price-Whelan, A. M., Brewer, J. M., Hogg, D. W., et al. 2018, ApJ, 854, 138}

\noindent
\footnotesize{Olander, T., Heiter, U., \& Kochukhov, O. 2021, A\&A, 649, A103}

\noindent
\footnotesize{Owen, J. E.,  \& Lai, D., 2018, MNRAS, 479, 5012}

\noindent
\footnotesize{Park, C., Jaffe, D. T., Yuk, I.-S., et al. 2014, Proc. SPIE, 9147, 91471D}

\noindent
\footnotesize{Pavlenko, Y. V. 2017, KPCB, 33, 55}

\noindent
\footnotesize{Perdigon, J., Laverny, P. de, Recio-Blanco, A., et al. 2021, A\&A, 647, A162}

\noindent
\footnotesize{Petigura, E. A., Marcy, G. W., Winn, J. N., et al. 2018, AJ, 155, 89}

\noindent
\footnotesize{Piaulet, C., Benneke, B., Rubenzahl, R. A., et al. 2021, AJ, 161, 70} 

\noindent
\footnotesize{Pinsonneault, M. H., DePoy, D. L., \& Coffee, M. 2001, ApJ, 556, L59}

\noindent
\footnotesize{Piskunov N. E., Kupka F., Ryabchikova T. A., Weiss W. W., Jeffery C. S. 1995, A\&AS, 112, 525}

\noindent
\footnotesize{Plez, B. 2012, Turbospectrum: Code for spectral synthesis}

\noindent
\footnotesize{Pollack, J. B., Hubickyj, O., Bodenheimer, P., et al. 1996, Icar, 124, 62}

\noindent
\footnotesize{Polanski, A. S., Crossfield, I. J. M, Howard, A.W., et al. 2022, Research Notes of the American Astronomical Society, 6, 155}

\noindent
\footnotesize{Putirka, K. D.,  \& Xu, S. 2021, NatCo, 12, 6168} 

\noindent
\footnotesize{Ram{\'i}rez, I., Khanal, S., Lichon, S. J., Chanam{\'e}, J., Endl, M., Mel{\'e}ndez, J., \& Lambert, D. L., 2019, MNRAS, 490, 2448}

\noindent
\footnotesize{Recio-Blanco, A., Laverny, P. De., Palicio, P. A., 2022, arXiv:2206.05541}

\noindent
\footnotesize{Reggiani, H.,  Schlaufman, K. C., Healy, B. F., et al. 2022, AJ, 163, 159}

\noindent
\footnotesize{Reiners, A., Zechmeister, M, Caballero, J. A., et al. 2018,A\&A, 612, A49}

\noindent
\footnotesize{Rothman, L. S., 2021, NatRP, 3, 302}

\noindent
\footnotesize{Ryabchikova, T., Piskunov, N., Kurucz, R. L. et al. 2015, Phys. Scr, 90, 054005}

\noindent
\footnotesize{Ryabchikova, T., Piskunov, N., \& Pakhomov, Y. 2022, arXiv:2209.13395} 

\noindent
\footnotesize{Santos, N. C., Israelian, G., \& Mayor, M. 2004, A\&A, 415, 1153}

\noindent
\footnotesize{Santos-Peral, P., Recio-Blanco, A., de Laverny, P., Fernández-Alvar, E., \& Ordenovic, C. 2020, A\&A, 639, A140}

\noindent
\footnotesize{Sawczynec, E., Mace, G.,  Gully-Santiago, M., et al. 2022, AAS, meeting No. 240, id.203.06}   

\noindent
\footnotesize{Schlaufman, K. C. 2015, ApJL, 799, L26}

\noindent
\footnotesize{Shan, Y.,   Reiners,  A.,  Fabbian, D., et al. 2021, A\&A,  654, A118}

\noindent
\footnotesize{Souto, D., Cunha, K., Smith, V., et al. 2016, ApJ, 830, 35}

\noindent
\footnotesize{Sneden, C., Lucatello, S., Ram, R. S., et al. 2014, ApJS, 214, 26}

\noindent
\footnotesize{Souto, D., Cunha, K., García-Hernández, D. A., et al. 2017, ApJ, 835, 239}

\noindent
\footnotesize{Souto, D., Unterborn, C. T., Smith, V. V., et al. 2018, ApJL, 860, L15}

\noindent
\footnotesize{Souto, D., Cunha, K., \& Smith, V. V. 2021, ApJ, 917, 11}

\noindent
\footnotesize{Souto, D., Cunha, K., Smith, V. V., et al. 2022, ApJ, 927, 123}

\noindent
\footnotesize{Spake, J. J., Sing, D. K., Evans, T. M., et al. 2018, Natur, 557, 68}

\noindent
\footnotesize{Spina, L., Sharma, P., Mel{\'e}ndez, J. 2021, NatAs, 5, 1163}

\noindent
\footnotesize{Su\'{a}rez-Andr\'{e}s, L., Israelian, G., Gonz\'{a}lez Hern\'{a}ndez, J. I., et al. 2018, A\&A, 614, A84}

\noindent
\footnotesize{Szab\'{o}, Gy. M.,  \& Kiss, L. L. 2011, ApJL, 727, L44} 

\noindent
\footnotesize{Teske, J. K., Thorngren, D., Fortney, J. J., et al. 2019, AJ, 158, 239}

\noindent
\footnotesize{Thiabaud, A. Marboeuf, U., Alibert, Y, et al. 2015, 580, A30}

\noindent
\footnotesize{Tsai, S-M., Lee, E. K. H., Powell, D., et al. 2022,  arXiv:2211.10490}

\noindent
\footnotesize{Tsuji, T., \&  Nakajima, T.  2014, PASJ, 66, 98}

\noindent
\footnotesize{Vissapragada, S.,  Knutson, H. A., Greklek-McKeon, M., et al. 2022, 164, 234}

\noindent
\footnotesize{Wang, J., \& Fischer, D. A. 2015, AJ, 149, 14}

\noindent
\footnotesize{Wilson, R. F., Teske, J., Majewski, S. R., et al. 2018, AJ, 155, 68}

\noindent
\footnotesize{Winn, J. N., Sanchis-Ojeda, R., Rogers, L., et al. 2017, AJ, 154, 60}

\noindent
\footnotesize{Woolf, V. M \& Wallerstein, G. 2020, MNRAS 494, 2718}

\noindent
\footnotesize{Xu, S. \&  Bonsor, A. 2021, Eleme, 17, 241}

\noindent
\footnotesize{Yuk, I.-S., Jaffe, D. T., Barnes, S., et al. 2010, Proc. SPIE, 7735, 77351M}

\noindent
\footnotesize{Zhu, W., Wang, J., \& Huang, C. 2016, ApJ, 832, 196}

\end{document}